\documentclass[a4paper,12pt]{article}
\usepackage[affil-it]{authblk}
\usepackage{cite}
\usepackage{amsmath}
\usepackage{graphicx}
\usepackage{xcolor}
\usepackage[cp1252]{inputenc}
\usepackage{float}
\usepackage{mathtools}
\usepackage{tabu}
\usepackage{amssymb}
\usepackage{tikz}

\providecommand{\keywords}[1]
{
  \small	
  \textbf{\textit{Keywords:}} #1
}
\usepackage{tikz}
\makeatletter
\DeclareRobustCommand*{\bfseries}{%
   \not@math@alphabet\bfseries\mathbf
   \fontseries\bfdefault\selectfont
   \boldmath
}
\makeatother
\usepackage[square,numbers,sort&compress]{natbib}
\begin{document}
\title{ On the  Existence of Logarithmic Terms in the Drag Coefficient and   Nusselt Number of a Single Sphere at High Reynolds Numbers }
\author[1,2]{Yousef M. F. El Hasadi\footnote{Email address for correspondence: G.DamianidisAlChasanti@tudelft.nl,yme0001@auburn.edu}}
\author[1]{ Johan T. Padding\footnote{Email address for correspondence:J.T.Padding@tudelft.nl} }
\affil[1]{Process and Energy department, Delft University of Technology,Leeghwaterstraat 39,2628 CB Delft, the Netherlands}
\affil[2] {Civil Engineering and Geosciences Department,Delft University of Technology, Stevinweg 1, 2628 CN Delft, the Netherlands }
\date{}
\maketitle
\begin{abstract}
At the beginning of the second half of the twentieth century, Proudman and Pearson (J. Fluid. Mech.,2(3), 1956, pp.237-262) suggested that the functional form of the drag coefficient ($C_D$) of a single sphere subjected to uniform fluid flow consists of a series of logarithmic and power terms of the Reynolds number ($Re$).\ In this paper, we will explore the validity of the above statement for Reynolds numbers up to $10^{6}$ by using a symbolic regression machine learning method.\ The algorithm is trained by available experimental data and data from well-known correlations from the literature for $Re$ ranging from $0.1$ to $2\times 10^5$.\ Our results show that the functional form of the $C_D$ contains powers of $\log(Re)$, plus the Stokes term, fulfilling partially the statement made above. The logarithmic $C_D$ expressions can generalize (extrapolate) beyond the training data and are the first in the literature to predict with acceptable accuracy the rapid decrease (drag crisis) of the $C_D$ at high $Re$.\ We also find a connection between the root of the $Re$-dependent terms in the $C_D$ expression and the first point of laminar separation.\ We did the same analysis for the problem of heat transfer under forced convection around a sphere and found that the logarithmic terms of $Re$ and Peclect number $Pe$ play an essential role in the variation of the  Nusselt number $Nu$.\ The machine learning algorithm independently found the asymptotic solution of Acrivos and Goddard (J. Fluid. Mech., 23(2),1965, pp.273-291).


\end{abstract}
\\

\section{Introduction}
Predicting the drag force on an object fixed in a planar flow has been the  subject of extensive investigation from the early days  of fluid mechanics when it emerged as an independent discipline.\ The analytical solution for the drag force experienced by a  rigid sphere for  creeping flow conditions, found by Stokes \cite{stokes1851effect} in 1851, is one of  the first known analytical expression in the fluid mechanic's community. Stokes assumed in his solution that  inertial effects of the fluid could be neglected throughout the solution domain.\ However, Oseen 
\cite{oseen1910uber} found an inconsistency in the Stokes solution. Specifically, he found that inertial fluid effects cannot be neglected far away from the sphere. \ He derived a new form of  equations, known as Oseen equations \cite{oseen1910uber},  that can handle this inconsistency, and he came up with an improved approximation  for the drag coefficient, defined as  $C_D = F_D/(\frac12 \rho v_{\infty}^2 \frac{\pi}{4} d^2)$,
where $F_D$ is the drag force, $\rho$ the fluid density, $v_{\infty}$ the fluid flow velocity far away from the sphere, and $d$ the sphere diameter
 \cite{oseen1913ueber}.\ There are additional solutions to the  Oseen equations,  such as those of Goldstein 
\cite{goldstein1929steady} and Fax\'en 
\cite{faxen1923bewegung}.\ 

 Proudman and Pearson\cite{proudman1957expansions} and Kaplun and Lagerstrom \cite{kaplun1957asymptotic} used the matched asymptotic method to solve the Navier-Stokes equations to resolve the fluid flow around different blunt bodies.
\ Proudman and Pearson \cite{proudman1957expansions} divided the flow field around the sphere into two stream function expansions.\ The first one, which they called the  Stokes expansion,  controls the flow near the surface of the sphere.\ The  second expansion, which they called  the  Oseen expansion,  controls the flow far from the surface of the sphere.\ Both expansions are based on the Navier-Stokes equations, and the two expansions are matched at a certain distance from the sphere using the method of matched asymptotics.\ Evaluating stresses from the Stokes expansion they arrived at  the following expression for the $C_D$ of a sphere: 
\begin{equation} \label{eq1}
C_D= \frac{24}{Re}(1+\frac{3}{16}Re+\frac{9}{160}Re^2\log(\frac {Re}{2}))
\end{equation}
 Here $Re = \rho v_{\infty} d/\mu$  is the Reynolds number.\ They  made the following statement (conjecture) about the expansions that govern the flow field
  \cite{proudman1957expansions}: ``\textit{The non-linearity of the Navier-Stokes equation then shows that both expansions must involve powers of $\log(Re)$, and it seems reasonable to suppose that both expansions are in powers of $Re$, each term of which is multiplied by polynomial in $\log(Re)$}''.\ This statement also reflects on the functional form of the drag coefficient.\ However, the authors did not mention the $Re$ range for which the statement is valid.\ From now on, we  will call this conjecture   \textbf{P\&P}. 
   Graebel  \cite{graebel2007advanced} supported the \textbf{P\&P} statement by mentioning that the $C_D$ functional form that will result from  asymptotic expansions of the Navier-Stokes equations will always be a function of $\log(Re)$.\ A few years later, Chester et al.\cite{chester1969flow}  added an extra term to Eq.(\ref{eq1}), which was the last addition that came from the expansion of the Navier-Stokes equations.\\
 
 { The appearance of  logarithmic terms (alternatively  known as logarithmic switchback 
  terms \cite{lagerstrom1984note}) in the asymptotic expansions have intrigued the scientific community, because in some instances  they were not   forced by the  governing equations \cite{popovic2005geometric} .\ 
  Van Dyke \cite{van1975perturbation} dedicated  a section in his book describing the proliferation of logarithmic terms in different fluid mechanics problems, and he made the following comment: ``\textit{one can philosophize that description by fractional powers fails to exhaust the myriad phenomena in the universe, and logarithms are the next simplest function}".\ Initially,   the logarithms were tied with paradoxes in  fluid mechanics, or to  the singular perturbation techniques themselves.\ However, Lagerstrom and Reinelt \cite{lagerstrom1984note}  showed that  logarithmic terms are part of the  solution of the governing equations, and the asymptotic expansion  method is just one way to reach to the solution.\ This view is supported by other investigations using different mathematical methods \cite{holzer2014analysis, popovic2004rigorous}.}\\

\ There are  analytical solutions for the Stokes and Oseen regimes for some non-spherical particles such as oblate or prolate  spheroids, circular cylinders  and few other particle geometries 
 \cite{happel2012low,cox1965steady,breach1961slow,
 aoi1955steady}.\ Eq.(\ref{eq1}) and all  other analytical solutions, regardless of the  shape of the particles,  are valid up to  $Re\approx$ 1.0.\ \\
 
 For higher $Re$, analytical solutions for the Navier-Stokes equations cease to exist due to its non-linearity.\ The flow around a sphere at high $Re$ consists of a mosaic of different flow morphologies, depending on  $Re$ as described by Achenbach \cite{achenbach1972experiments}, and  Kambel and Girimaji \cite{kamble2020characterization}.\ High Reynolds number flows ($Re\geq 10^4$) are usually classified into four flow regimes.\ In the subcritical flow regime, the $C_D$ value is independent of $Re$. In contrast, in the critical flow regime,  $C_D$ starts to decrease rapidly as  $Re$ increases until a minimum is reached at a critical Reynolds number.\ For a smooth sphere  $Re_{cr} \approx 3.7\times 10^5$.\ This critical flow regime some times is referred to as the drag crisis.\ Beyond the critical $Re$, in the so-called supercritical regime, the drag coefficient slowly increases with increasing $Re$ until it reaches a maximum value.\ Further increasing  $Re$, the drag coefficient stays constant and this regime is called transcritical. For  the prediction of $C_D$ at high $Re$ one usually resorts to numerical simulations \cite{dennis1971calculation,jenson1959viscous,
 nakhostin2019investigation,constantinescu2002detached} or experiments \cite{achenbach1972experiments,deshpande2017intermittency,maxworthy1969experiments}.
The results of these numerical simulations and experiments are translated into  fitting correlations,  with a range of applicability limited to the range of the data that is used in the fitting process.\ This has resulted in a zoo of  correlations that take different mathematical  forms  \cite{rouse1961fluid,engelund1967monograph,clift1970motion,
morsi1972investigation,graf1984hydraulics,flemmer1986drag,
khan1987resistance,swamee1991drag}, as shown  in the  extensive list published in the recent review by Goossens \cite {goossens2019review}. \ The majority of  correlations focus on the subcritical regime and  take the following functional form: 
\begin{equation} \label{eq2}
C_D=\underbrace{ \underbrace{\frac{24}{Re}(C_1+C_2Re^a)}_{\text{Schiller and Naumman}}+\frac{C_3}{1+\frac{C_4}{Re}}}_{\text{Brown and Lawler}}
\end{equation}
 The second term of  Eq.(\ref{eq2}) arises from  boundary layer theory \cite{schlichting2016boundary}, which accounts for the inertial effects of the fluid.\ The value of the  exponent $a$  ranges from  0.5 to 0.68.\ These type of correlations are suitable for $Re$  up to  $2\times 10^5$, right before  drag-crisis. \\
 
Concerning  the heat transfer rate from a particle fixed in a fluid, most  investigations available in the literature are related to the case of  forced convection.\ In this type of flow, the velocity profile is decoupled from that of the temperature. \ For further simplification, there is also no variation in the transport properties of the fluid with temperature.\ These simplifications pave the way of obtaining several analytical solutions for a single sphere \cite{acrivos1962heat} for   limited cases of low $Re$ and Peclet number $Pe = v_{\infty} d / \alpha$, where $\alpha$ is the thermal diffusivity of the fluid.\ Acrivos and Taylor \cite{acrivos1962heat}  used  asymptotic expansions  and the  velocity profile of the Stokes solution to find  the following relation for the  Nusselt number $Nu = hd/k$, where $h$ is the (convective and surface mean) heat transfer coefficient and $k$ is the thermal conductivity of the fluid (linked to the thermal diffusivity through $k = \alpha \rho c_p$, with $c_p$ the specific heat capacity of the fluid), 
 for the case of $Pe\to$ 0 and $Re\to$ 0 : 
\begin{equation}\label{eq3}
Nu = 2+\frac{1}{2}Pe+\frac{1}{4}Pe^2\log(Pe)+0.034 Pe^2 +\frac{1}{16}Pe^2\log(Pe)
\end{equation}
In practice, this solution is limited to $Re \lesssim$ 0.03.\ Rimmer \cite{rimmer1968heat} added an extra term to Eq.(\ref{eq3}) from asymptotic expansions, and as far as we know this is the last term that evolved from the matched asymptomatic expansions in the low $Pe$ and  $Re \to 0$ regime. \ Conversely, for   
  $Pe\to \infty$ and  $Re \to 0$,  Acrivos and Goddard \cite{acrivos1965asymptotic} used the matched asymptotic expansions  to  arrive at the following relation for  $Nu$: 
\begin{equation}\label{eq4}
Nu =0.922+ 1.249Pe^{\frac{1}{3}}
\end{equation}

As for the case of drag, for higher $Re$ we need to rely on semi-empirical relations to express the variation of  $Nu$ with the flow field parameters.\ Whitaker\cite{whitaker1972forced} provided a correlation, which is still considered one of the most accurate available in  literature \cite{sparrow2004archival}: 
\begin{equation} \label{eq5}
Nu =  2+ (C_4Re^{a_1}+C_5Re^{a_2})Pr^{a_3}
\end{equation}
where $Pr = c_p \mu / k$ is the Prandtl number (note that $Pe = Re Pr$).\ The values of $a_1$, $a_2$, and $a_3$ are $\frac{1}{2}$, $\frac{2}{3}$, and 0.4, respectively.\ The Whitaker correlation is valid for 1 $\leqslant Re \leqslant 10^5$ and a wide range of $Pr$.\ The second, and  third terms represent  inertial fluid effects, and their functional form is inspired by  boundary layer theory.\ Although the first term comes from the analytical solution for  pure conduction from a sphere, all exponents in Eq.(\ref{eq5}) are obtained from empirical fitting.\\

In summary, almost all correlations for drag and heat transfer found in literature are expressed as power law expansions, similar to Eqs. (2), (4) and (5). Correlations with logarithmic terms, such as Eqs. (1) and (3), are extremely rare and seem to have been largely overlooked.\\

The improvement of high-performance computer architectures, plus the availability of data from numerical simulations and experiments, sparked an increase  in interest to use machine learning methods to solve  problems in many scientific disciplines.\ This has led to label machine learning as the fourth paradigm in science, next to experimentation, theory and simulation \cite{butler2018machine}.\ When it comes to fluid mechanics, applying machine learning methods constitutes a challenge for several reasons, such as the transient nature of most fluid mechanics problems, the heterogeneity of most available data, the extensive non-linearities that govern fluid mechanics, and the multi-scale nature  of most problems in hand \cite{brunton2020machine}.\ To deal with these challenges, an  ideal machine learning algorithm for fluid mechanics, should possess features such as interpretability, explainability, generalisability, and convergence \cite{brunton2020machine}.\ One of the most popular machine learning frameworks that are used extensively in different fluid mechanics problems, from solving partial differential equations \cite{dissanayake1994neural,raissi2018hidden}, discovering physics \cite{iten2020discovering}, learning the active-nematic hydrodynamics \cite{colen2020machine}, and predicting  physical properties \cite{kushvaha2020artificial} are the artificial neural networks (ANN).\ Other machine learning methods that are used for scientific discovery are  sparse identification of nonlinear dynamical systems for discovering differential equations from sparse data \cite{brunton2016discovering}, and symbolic regression that is used for discovering laws of nature \cite{schmidt2009distilling}, discovering new materials \cite{weng2020simple}, and solving 
fluid flow problems \cite{el2019solving}.   \\
In this paper we will use symbolic regression, which is a modern tool for unbiased determination of correlations, to re-investigate known data on drag and heat transfer.\ We will show that symbolic regression actually rediscovers the logarithmic terms, suggesting that logarithmic expansions may represent the physics better than power law expansions. As a side result, we will show that there is an intriguing connection between the found logarithmic terms and the point of first boundary layer separation.
 
\section{Methodology} 
In this paper, we will use the symbolic regression machine learning method proposed by Koza \cite{koza1992genetic}.\ Symbolic regression is a powerful tool for searching the mathematical space for an approximate functional relation between a certain number of input and output variables, and it is based on genetic programming proposed by 
Holland \cite{holland1992adaptation}.\ The framework of genetic programming is probabilistic, and is not based on mathematical principles, such as correctness, consistency, justifiability, certainty, orderliness, and decisiveness as outlined by Koza \cite{koza1992genetic}, but solely on the principles of Darwinian evolution \cite{darwin2004origin}.\ The idea of the genetic programming is simple, and it is based on transforming an initial population (in our case a population of  mathematical functions) to a new population that survived a particular fitness constraint.\ The main operators that are used to create the new population are similar to those found in nature, namely that of reproduction and crossover 
 \cite{koza1992genetic} .\\

The algorithm first generates a random pool of functions, that undergo genetic operations such as crossover, which corresponds to the combination of two functions  to give a new offspring function.\ Another operation is a mutation in which a certain part of the mathematical function is changed randomly.\  Two indexes measure the fitness of the newly obtained functions.    The first index  is minimizing the mean square difference between the training and predicted dependent values.\ The  second index is to check the mathematical complexity of functions, and select the simplest ones, to prevent over-fitting.\ We used the Eureqa software \cite{schmidt2009distilling}  as symbolic regression    platform.\ A  rigorous description of the symbolic regression algorithm  in use in the current investigation  is given in \cite{el2019solving}.\\

In  Appendix A we illustrate that the machine learning algorithm we use can capture a known function's series expansion.\ It shows the ability of symbolic regression to find expansions of functions, that are valid beyond the training data used to obtain them, which gives symbolic regression an advantage compared to, for instance, artificial neural networks.

\section{Results}
In the first subsection, we will explore the  dependence of the drag coefficient  $C_D$ on $Re$ for a fixed sphere.\ We will devote the second subsection to explore the  dependence of the Nusselt number   $Nu$ of a sphere on $Re$ and $Pe$ (or $Pr$) for the case of forced convection with constant transport properties.  

\subsection{Drag coefficient $C_D$ }
We will start by exploring the $C_D$ dependency on $Re$ for the case of a sphere.\ We will create three data sets for the regression process. The first one will be generated from the correlation of Brown and Lawler \cite{brown2003sphere} which has the functional  shape of Eq.(\ref{eq2}).\ This data set contains about 8500 points in the range $ 0.1 < Re < 1.9\times 10^5$, which is enough to capture the smallest details in the $C_D$ variation.\ The second data set that we will use is the exact experimental data  that Brown and Lawler \cite{brown2003sphere} used themselves  to derive their correlation.\ It contains about 450 points in the range $0.1 < Re < 1.975\times 10^5$ .\ The final data set is based on the Schilller and Naumann \cite{schiller1933grundlegenden} correlation, and contains  of 5020 points in the range $0.1 < Re < 700$. \\ 

\ We will start by examining the first data set, and  we will  let the symbolic regression  algorithm guess about the functional form of the $C_D$ dependence on $Re$.\ We can do this by specifying the most general initial functional form: 

\begin{equation} \label{eq6}
C_D =f(Re)
\end{equation}
The algorithm derived several regression equations, but here we will show  two, one because it accurately fits the results, and the other  because it is simple.\ The equations are the following: 
\begin{equation} \label{eq7}
C_D = a_1 + \frac{a_2}{Re} + a_3\sqrt{Re} + \frac{a_4}{\sqrt{Re}}+ \frac{a_5}{(a_6+Re)} +a_7Re
\end{equation}
\begin{equation}\label{eq8}
C_D=a_1 + \frac{a_2}{Re} + \frac{a_3}{\sqrt{Re}}
\end{equation}
The coefficients of Eq.(\ref{eq7}), and Eq.(\ref{eq8}) are listed in Table \ref{table1}.\ Eq.(\ref{eq8}) contains  the Stokes  $\frac{1}{Re}$ term, and the first-order term from  boundary layer theory $\frac{1}{\sqrt{Re}}$.\ The first known dependency of  $C_D$ on  $\frac{1}{\sqrt{Re}}$  came from the Blasius solution \cite{blasius1908grenzschichten}  of the boundary layer equations proposed by Prandtl \cite{Prandtl1904} for the case of a flat plate.\ The $C_D$ for blunt bodies, like a sphere, has a similar dependency on $Re$  \cite{leal2007advanced,abraham1970functional}.\  A similar form as Eq.(\ref{eq8}) was  obtained previously by fitting experimental data  \cite{brauer1972stromungswiderstand,holzer2008new}, and also by using concepts of boundary layer theory \cite{abraham1970functional}.\ 
Refs \cite{brauer1972stromungswiderstand,holzer2008new}  used 
non-linear fitting tools to obtain their correlations,  which require  \textit{a priori} knowledge of the functional structure.\ A comparison between the the coefficients of Eq.(\ref{eq8}), and those of  Refs     \cite{brauer1972stromungswiderstand,holzer2008new,abraham1970functional} is given in Table \ref{table2}.\ The coefficients of Eq.(\ref{eq8}) have similar values to those of \cite{brauer1972stromungswiderstand}.\ Compared to those of \cite{holzer2008new} there is only significant difference in the value of $a_3$.\ There is  also a significant difference between the coefficients of Eq.(\ref{eq8}) and those of Abraham \cite{abraham1970functional}.\ This may be due to the pure theoretical nature of the equation proposed by Abraham.\\

 It is important to note that both the Stokes term and the boundary layer term have been found without using any sophisticated mathematical approach.\ On the contrary, they have been found by a probabilistic genetic  algorithm.\ The emergence of the boundary layer term in Eqs. (\ref{eq7}) and (\ref{eq8}) without human intervention can be added to the experimental and numerical results that support boundary layer theory, even though there is no general mathematical proof of its existence, as mentioned by Batchelor \cite{batchelor2000introduction}.\\

We will now try to explore the existence of  logarithmic switchback terms for the drag on a sphere for the { higher $Re$ regime}.\ We will use for this the first data-set  (i.e. data from the Brown and Lawler \cite{brown2003sphere} correlation).\ We will start by imposing the following initial functional form:
\begin{equation}\label{eq9}
C_D = f(\frac{24}{Re}, \log(Re), Re\log(Re), \log^2(Re))
\end{equation} 
We choose this form of the initial function because we want to ensure that logarithmic switchback terms similar to Eq.(\ref{eq1}) will be part of the initial soup of functions that the symbolic algorithm will further evolve.\ The symbolic regression algorithm converged to the 
 following equation : 
  \begin{equation}\label{eq10}
	C_D = a_1 +{\frac{a_2}{Re}} +{a_3\log(Re)+ a_4\log^2(Re) + a_5\log^4(Re)} 
  \end{equation}
  
  The values of the coefficients of Eq.(\ref{eq10}) are listed in Table \ref{table3}. Eq.(\ref{eq10}) depends on  powers of $\log(Re)$ and also contains the  Stokes law term.\ The form of Eq.(\ref{eq10}) is partially fulfilling the  \textbf{P\&P} conjecture \cite{proudman1957expansions} for $Re$ as high as $2\times 10^{5}$.\ Overall, Proudman and Pearson \cite{proudman1957expansions} made a profound statement more than 64 years ago, using only  mathematical intuition, and they may have been right when they  suspected that  logarithmic switchback terms are part of the solution. 
   \ It may be difficult for the current form of the genetic algorithm to spot the entire logarithmic switchback  series,  because  reducing the complexity of the equations is part of its optimization process.\ Therefore,   terms that do not play a  significant role in the variation of the dependent variable ($C_D$) will die out during the evolution process.\  The failure of detection of  $Re^n\log^n(Re)$ terms,  where $n$ is an integer, after a  significant number of mathematical formula evaluations  exceeding $10^{11}$,  suggests that their signal is weak (a metaphor for their insignificant role in the dependence of $C_D$ on $Re$).\ If we read more carefully the conjecture, we  find that Proudman and Pearson \cite{proudman1957expansions} used the following wording:\textit{ `` It seems reasonable to suppose that both expansions are in powers of Re"}.  They used the word `reasonable to suppose', expressing doubt, while for the   $\log(Re)$ terms they used the word `must' which reflects that the authors were sure about their appearance in the two expansions.\ Adding to that, Proudman \cite{chester1969flow} was frustrated about the poor convergence  of his equation, mainly because it is only valid for extremely low values of $Re$.\ He suggested that the expansion in  powers of $Re$ may be a poor idea \cite{chester1969flow, hunter1990lagerstrom}.\\

\begin{figure}[p]
\begin{center}
\includegraphics [scale=0.8, trim = 0 300 0 0,clip]{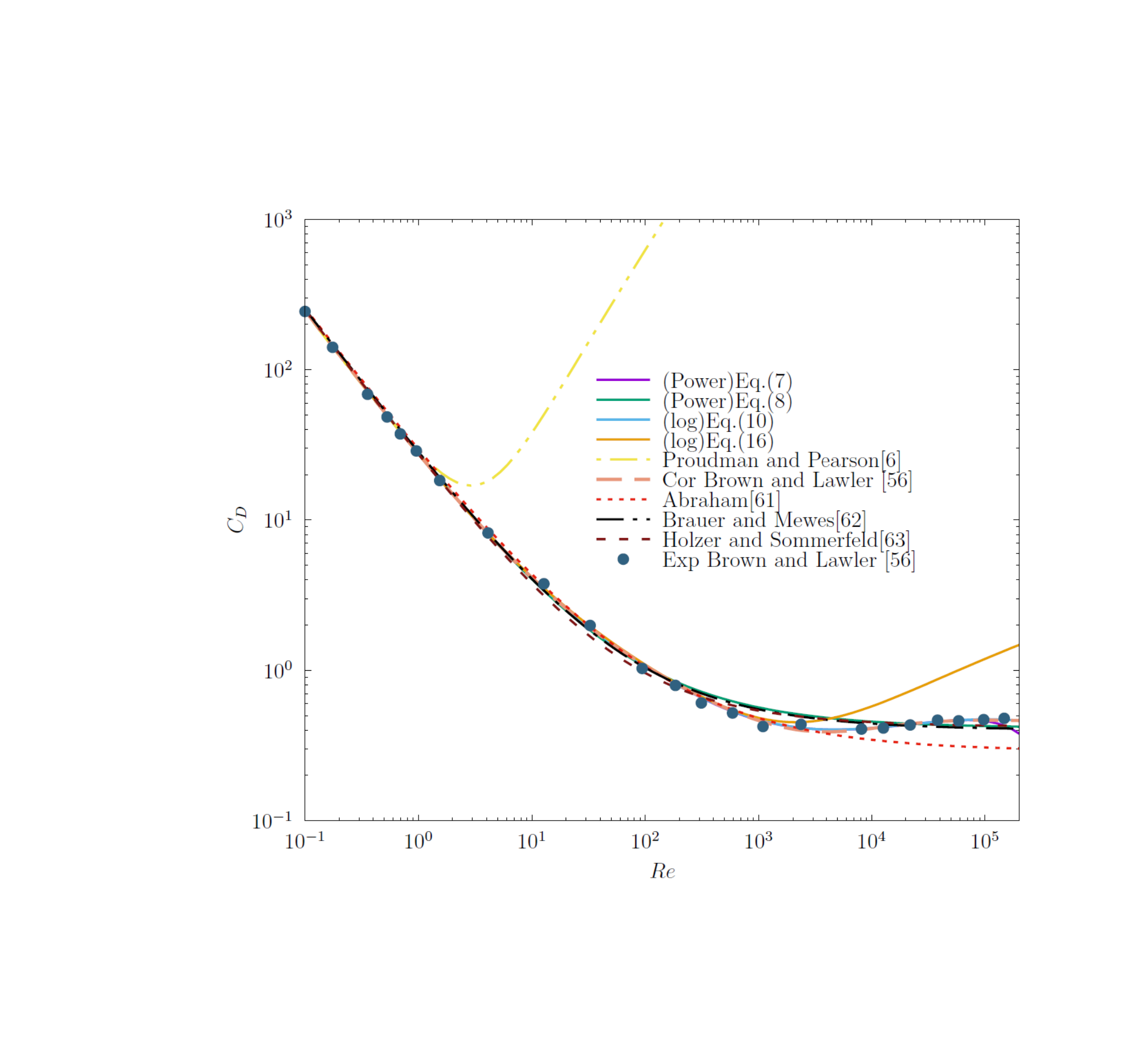}
\end{center}
\caption{Comparison  between the drag coefficient $C_D$ predicted by  Eq.(\ref{eq7}), Eq.(\ref{eq8}),  Eq.(\ref{eq10}), Eq.(\ref{eq16})and, different sources from the literature.\ Dashed lines indicate literature correlations. Symbols indicate experimental values. }
\label{Fig1}
\end{figure}

 In order to further validate the ecosystem of the equations that we obtained, we will compare their predictions with various sources in the literature, as shown in Figure \ref{Fig1}.\ The first insight from  Figure \ref{Fig1} is that  Eq.(\ref{eq1}) is valid only at low $Re$, and this was one of the main reasons we believe that the scientific community did not  further explore the use of  logarithmic terms, even as fitting functions.\ Eq.(\ref{eq7}) and Eq.(\ref{eq10}) follow closely the correlation of Brown and Lawler \cite{brown2003sphere}, and also the experimental data used to obtain  their correlation.\ The average relative errors  between the predictions of Eq.(\ref{eq7}) and Eq.(\ref{eq10})  with respect to the experimental results of \cite{brown2003sphere} are 3.87\% and 3.39\%, respectively.\ We see that Eq.(\ref{eq8}) follows closely the results of \cite{holzer2008new,brauer1972stromungswiderstand}, while it deviates from the predictions  of Abraham \cite{abraham1970functional} especially for values of $Re$  above $10^3$.\ This is expected because the equation provided by Abraham \cite{abraham1970functional} is valid for $Re$ up to $10^3$.\ Also, Eq.(\ref{eq8}) and those of 
 references \cite{brauer1972stromungswiderstand, holzer2008new, abraham1970functional} cannot capture the local minimum for Re between $10^3$ and $10^4$ that the experimental results of \cite{brown2003sphere} show.  \\

 \ Comparing   Eq.(\ref{eq7}) and Eq.(\ref{eq10}), we find that  their complexity  index is  34 and 19, respectively.\ The complexity index shows that the logarithmic series  representation of  $C_D$ is mathematically simpler compared to the power series representation, making Eq.(\ref{eq10}) more favourite to represent the physical phenomena of the $C_D$ variation  according to  Occam's razor statements \cite{domingos1999role}.\ One of these statements is: ``\textit{Given two models with the same generalization error, the simpler one should
be preferred because simplicity is desirable in itself.}"\\

\begin{figure}[p]
\begin{center}
\includegraphics [scale=0.8, trim = 0 300 0 0,clip]{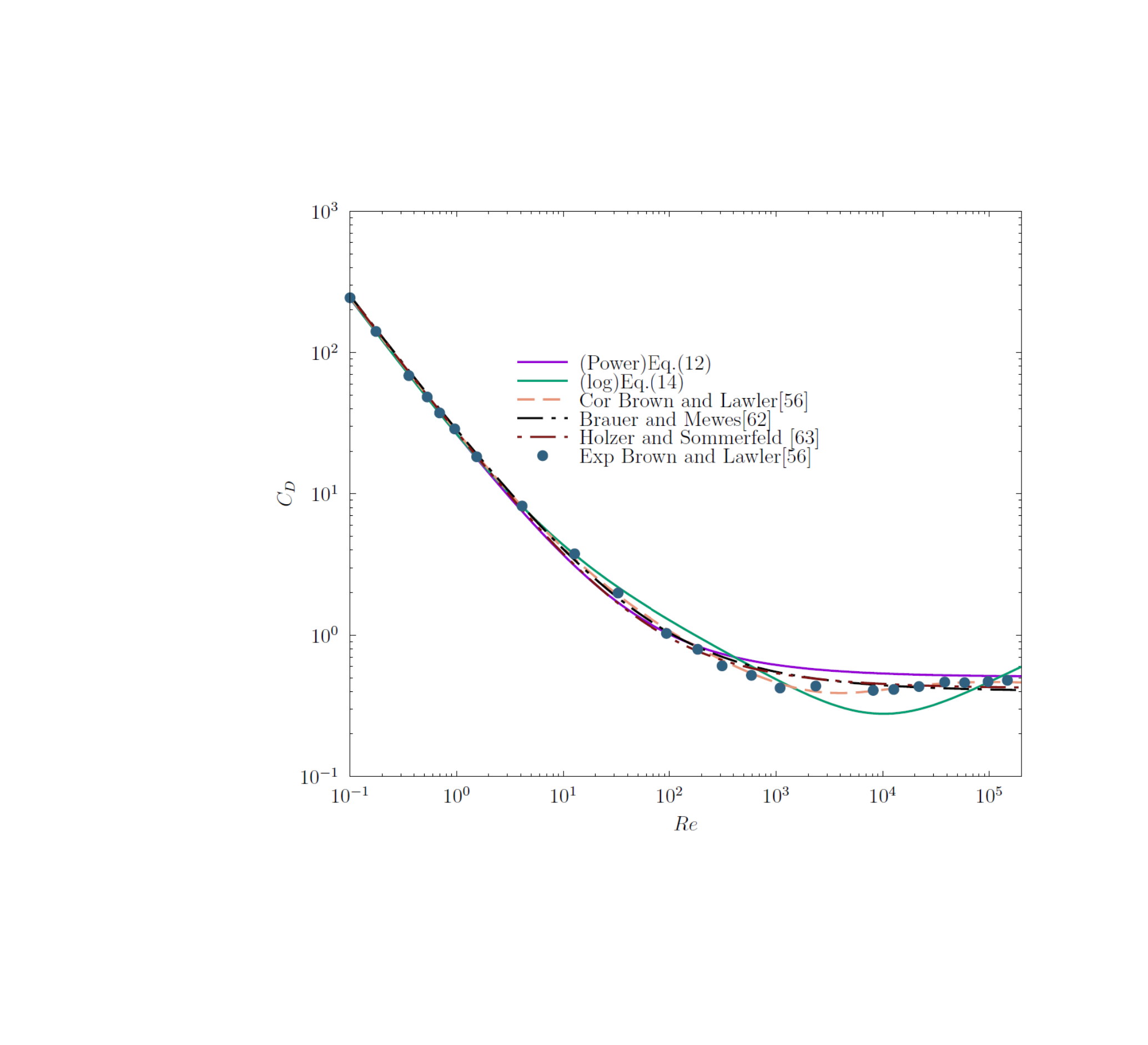}
\end{center}
\caption{Comparison  between drag coefficient $C_D$ predicted by  Eq.(\ref{eq12})  Eq.(\ref{eq14}), and different sources from the literature.\ Dashed lines indicate literature correlations. Symbols indicate experimental values. }
\label{Fig3}
\end{figure}

Now we will use the second (experimental) data set, to explore the feasibility of getting predictive equations for $C_D$ from a limited amount of noisy experimental data.\ We will start by letting the algorithm guess the $C_D$ dependence: 
\begin{equation} \label{eq11}
C_D =f(Re)
\end{equation} 
The symbolic regression algorithm found the following equation: 
\begin{equation} \label{eq12}
C_D = a_1 + \frac{a_2}{Re} + \frac{a_3}{\sqrt{Re}}
\end{equation}
The coefficients of Eq.(\ref{eq12}) are listed in Table \ref{table1}.\ Using the second data set we next  explore if the data show any  logarithmic dependence  by  imposing the following initial set of functions: 
\begin{equation} \label{eq13}
C_D = f(\frac{24}{Re}, \log(Re), Re\log(Re), \log^2(Re))
\end{equation}
We got the following equation for $C_D$: 
\begin{equation} \label{eq14}
C_D =  a_1 +\frac{a_2}{Re} +\frac{ a_3\log^2(Re)}{Re}+   a_4\log(Re)+ a_5\log^2(Re)
\end{equation}
The values of the coefficients are listed in Table \ref{table3}.\ Eq.(\ref{eq12}) is of a similar form as Eq.(\ref{eq8}), but the coefficients are not identical, because the second data set contains far less data, and also  contains some noise.\  The derivation of Eq.(\ref{eq12}) from pure experimental data, without imposing  knowledge of any physics, except the definition of $Re$, shows that the symbolic regression algorithm discovered the Stokes limit and the term attributed to  boundary layer theory  without any external help.\ The algorithm needed less than an hour to discover what took human intellect  hundreds of years to achieve.\ However, the human factor is still required  since we have to select the equations that we think  represent physical reality from the population of equations  that the algorithm suggests.\ Eq.(\ref{eq14}) shows that we can get the logarithmic dependence from a pure experimental data set, and  it partially fulfils the \textbf{P\&P} conjecture. Eq.(\ref{eq14}) and Eq.(\ref{eq10}) are quite similar.\ We believe that Eq.(\ref{eq14}) failed to capture the $\log^4(Re)$ term because this term influences  $C_D$  in the high $Re$ regime where  there are significant fluctuations in the experimental  data set. \ Probably if there were a higher volume of data, especially at higher $Re$, the $\log^4(Re)$ term could also be captured from pure experimental results.\ A comparison of the  performance of the power expansion  Eq.(\ref{eq12}) and the logarithmic expansion Eq.(\ref{eq14}) against existing data in the literature is shown in Figure. \ref{Fig3}.\ The average relative error for Eq.(\ref{eq12}) and Eq.(\ref{eq14}) is 13.7\% and 12.0\%, respectively, against the experimental results of \cite{brown2003sphere}.\  Eq.(\ref{eq14}) shows a local minimum in the range of the $Re$ close to that of the experimental results of  \cite{brown2003sphere}, while Eq.(\ref{eq12}) fails to show any local minimum. \\

 {We will use the third and final data set from the Schiller and Naumann \cite{schiller1933grundlegenden} correlation  which contains information about the variation of $C_D$  for $Re$ ranging from 0.1 to 700. \ We will use the following general initial functional form:
\begin{equation}\label{15extra}
C_D = f(Re)
\end{equation}
The symbolic regression algorithm found the following equation for  $C_D$: 
\begin{equation}\label{16extra}
C_D =  a_1 + \frac{a_2}{Re} +a_3\log(Re)+ a_4\log^2(Re) 
\end{equation}

The   coefficients of Eq.(\ref{16extra}) are listed in Table \ref{table3extra}.\ The genetic algorithm came up with the logarithmic dependence of $C_D$ on $Re$  without any external help, and it discovered the \textbf{P\& P} conjecture partially.\ The value of $a_1$ = 3.1406, differs from the value of $\pi$  by only about 0.03\%.\ It will be very interesting in the future to investigate the value of $a_1$ by fitting to very accurate numerical or experimental data.\  Eq.(\ref{16extra}) follows the Brown and Lawler correlation \cite{brown2003sphere} up to  $Re$ of $10^3$, as shown in Figure \ref{Fig1}.\ This behaviour is expected because  higher power  logarithmic terms  are missing from Eq.(\ref{16extra}), since the training data was limited to $Re$ up to 700.}\\

\begin{figure}[p]
\centering
\includegraphics [scale=0.8, trim = 0 300 0 0,clip]{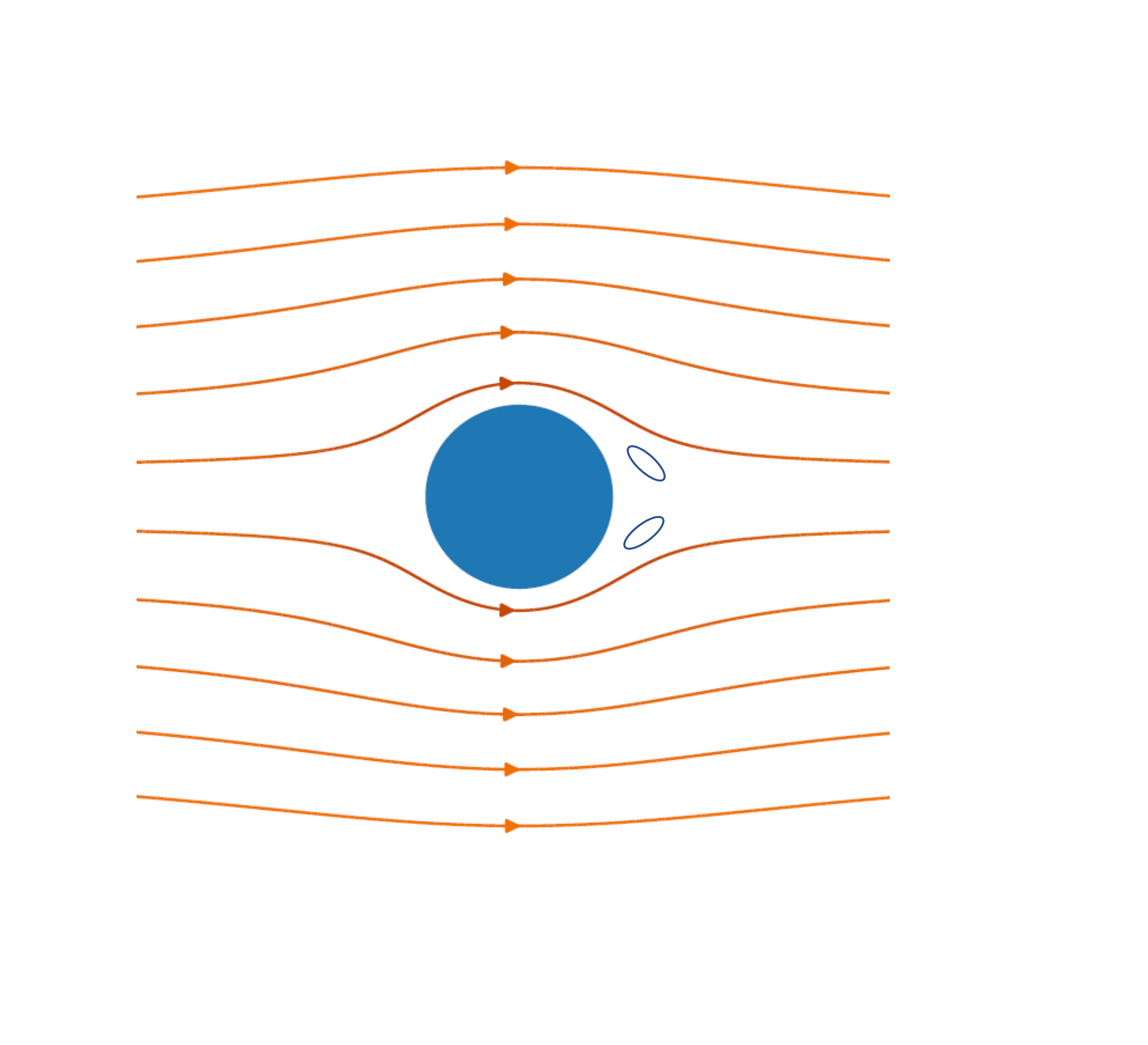}
\caption{Schematic of separated flow around a sphere}
\label{FigST}
\end{figure}

Up to this point we have discussed the drag without referring to the flow around the sphere.\ The flow around a sphere is a rich mosaic of phenomena, and usually drag correlations, fail to predict them.\ Among these  phenomena is the  emergence of a laminar separation point, which  is well known to occur for sufficiently blunt  objects, including a sphere.\ The point of  laminar separation is identified by the formation of a closed recirculating ring eddy at the rear of the sphere, as indicated in Figure \ref{FigST}.\ The first emergence of separation  is difficult to detect either experimentally or theoretically.\ For this reason, there is some discrepancy  in the literature on the value of the reported critical $Re_s$, and  corresponding drag $C_{Ds}$, at first separation.\ The first experimental observations by Nisi and Porter \cite{nisi1923lxxxvi} suggested  that $Re_s$ = 10.\ This was  confirmed by numerical simulations of Rimon and Cheng \cite{ rimon1969numerical}. \ On the other hand, Proudman and Pearson \cite{proudman1957expansions}, and  Van Dyke \cite{van1975perturbation}, by using the Stokes second  expansion,   estimated that $Re_s$ = 16, close to the numerical results of Bourot \cite{bourot1969comptes} and Jenson \cite{jenson1959viscous} of  15.2 and 17, respectively, and the experiments of Payard and Countanceau \cite{payard1974etude} indicating  $Re_s$ = 17.\ Other  simulation results    \cite{dennis1971calculation,chang1994unsteady}     show that $Re_s$ is equal to approximately 20, and the experiments of Taneda \cite{taneda1956experimental} predict that $Re_s$ = 24.\\
 
 \ If we inspect $a_1$ of the logarithmic expansion Eq.(\ref{eq10}) in Table \ref{table3} we  see that its value is  3.286, which is quite similar to the value of the drag coefficient $C_{Ds}$ at the initial laminar separation reported by \cite{payard1974etude}, which  is 3.306.\ If the constant $a_1$  is the drag coefficient at initial laminar separation, then the following transcendental equation must have a positive root at the corresponding Reynolds number $Re_s$:  
\begin{equation}\label{eq15}
\frac{a_2}{Re} + a_3\log(Re)+ a_4\log^2(Re) + a_5\log^4(Re)=0
\end{equation}    
By solving Eq.(\ref{eq15}) we find that  $Re_{rt}$ = 14.06 is its only root.\ That makes  $Re_{rt}$  the only $Re$ value that zeroes off all terms beyond the constant $a_1$.\ This  $Re_{rt}$  is close to values of  $Re_s$ reported in literature.\ For example, the relative error with respect to the results of Bourot\cite{bourot1969comptes} and Chang and Maxey \cite{chang1994unsteady} is 8\% and 30\%, respectively.\ We conjecture that $Re_{rt}$ is representing $Re_s$, even though   we do not have any  proof for this.\ We believe we are witnessing an instance where the machine learning algorithm found a mathematical description of a physical phenomenon, which needs human  abilities to be interpreted  in terms  of  physical laws.\ Otherwise, it will be a good approximation, that can describe some of the physics involved in the process of flow separation.\ As far as the authors are aware, there is  only one analytical prediction for the  point of first flow separation, from  slow motion viscous theory \cite{proudman1957expansions,van1970extension}.\ However, that result was disputed  by the authors of   \cite{proudman1957expansions,van1970extension}, as we will show later.   In practice, we depend on  numerical simulations to find the point of zero local shear stress, as described by  boundary layer theory \cite{schlichting2016boundary}. However, Batchelor \cite{batchelor2000introduction} raised serious doubts about estimating the onset of  separation by this method.\\

\ Beyond this point, we will assume that  (the smallest, real) root $Re_{rt}$ is equal to $Re_s$.\ Using the same procedure to calculate  $Re_s$,  from the logarithmic  Eq.(\ref{eq14}) by solving the following  transcendental equation:
\begin{equation}\label{eq16}
\frac{a_2}{Re} +\frac{ a_3\log^2(Re)}{Re}+   a_4\log(Re)+ a_5\log^2(Re)=0
\end{equation}
 we found the two following roots: $Re_s$ =  15.76, and $9.52\times 10^7$.\ The large root value of $ 9.52\times 10^7$,  is a non-physical result, which we believe is caused by the missing higher power $\log(Re)$ term from Eq.(\ref{eq14}).\ However, $Re_s$ = 15.76 compares very well with the results  of Bourot\cite{bourot1969comptes} and Chang and Maxey \cite{chang1994unsteady}, with a relative difference of 3.68\% and 21.2\%, respectively.\ If we do the same analysis for the logarithmic  Eq.(\ref{16extra}), we will find that $Re_s$ =  15.19, and $3.518\times 10^6$.\ For the smallest root,   the relative difference with the results of Bourot\cite{bourot1969comptes} and  Chang and Maxey \cite{chang1994unsteady} is 0.13\%, and 24.0\%, respectively. \\ 
  
We will  next calculate $Re_{rt}$ from the more popular power-law expressions  Eq.(\ref{eq7}) and Eq.(\ref{eq8}) in  the same way.\ For Eq.(\ref{eq7}) we find the following roots $Re_{rt1} = -2461-767i$, $ Re_{rt2} =-2461+767i$, and $Re_{rt3} = 3\times 10^5$.\ The first two roots are non-physical, while the third root, closely approximates the critical Reynolds number ($Re_{cr}\thickapprox  3.7\times 10^5$)  for the critical flow regime (drag crisis) as reported by Achenbach \cite{achenbach1972experiments}.\ We will further discuss the physical significance of   $Re_{rt3}$ in the generalization subsection since the value of$Re_{rt3}$  is outside the training data range. \ As for power-law Eq.(\ref{eq8}), it does not have any roots, neither in the real nor in the complex domain.\\

Returning to the logarithmic ecosystem of equations,  in their seminal works, Proudman and 
Pearson \cite{proudman1957expansions} and  Van Dyke \cite{van1975perturbation}  calculated the $Re_s$ value to be  16 analytically from the first and second terms in the Stokes expansion. Proudman 
and Pearson\cite{proudman1957expansions} made the following comment:  \textit{``This Reynolds number is far too large to make estimates based on only two terms of the Stokes expansion at all reliable.\ In fact, it cannot seriously be claimed that slow-motion theory gives even a qualitative expansion of the phenomena.''}\ However, Van Dyke \cite{van1975perturbation}  and Ranger\cite{ranger1972applicability} tried to confirm the result of Proudman and Pearson\cite{proudman1957expansions}, by using  extra terms in the Stokes expansion that contain the logarithmic terms from the results of Proudman and Pearson \cite{proudman1957expansions} and those of Chester et al. \cite{chester1969flow}.\ They failed because the Stokes expansion equation that includes the logarithmic terms has only complex roots.\ Van Dyke \cite{van1975perturbation} commented on this issue saying that\textit{ `` the logarithm needs reinterpretation."}\ In our work we now see that  the values of $Re_s$ from Eq.(\ref{eq10}), Eq.(\ref{eq14}), and Eq.(\ref{16extra}) are converging with different degree of accuracy toward a value of approximately 16.\\

.   \\
\subsection{Generalization beyond the training data}
In this subsection, we will  test our newly derived equations generalisation behaviour, for flow regimes that were not  included in the training data.\ Specifically, we will test their behaviour for the low Reynolds number regime for $Re$ down to $10^{-4}$, and for the critical flow regime for $Re$ up to $10^6$. 
\subsubsection{Low $Re$ flow regime}
In the low $Re$ regime, $\dfrac{24}{Re}$ is the dominant term for the drag coefficient, which will make it difficult to assess the performance of our equations, against the existing correlations, analytical solutions, experimental and numerical results.\ For this reason, we will use the way Maxworthy \cite{maxworthy1965accurate} plotted his drag coefficient data.\ He plotted the quantity $\dfrac{C_D}{C_{Ds}}-1$ against  $Re$,where $C_{Ds}$ is  the Stokes drag ($\dfrac{24}{Re}$). This way, we eliminate the divergence of the Stokes term, which makes the comparison with different sources from the literature more precise.\ From  low Reynolds number theory we know that $\dfrac{C_D}{C_{Ds}}-1$ converges  to $\dfrac{3}{16}Re$  (Oseen term) for extremely  low $Re$.\\

\begin{figure}[p]
\begin{center}
 \includegraphics [scale=0.8, trim = 0 250 0 0,clip]{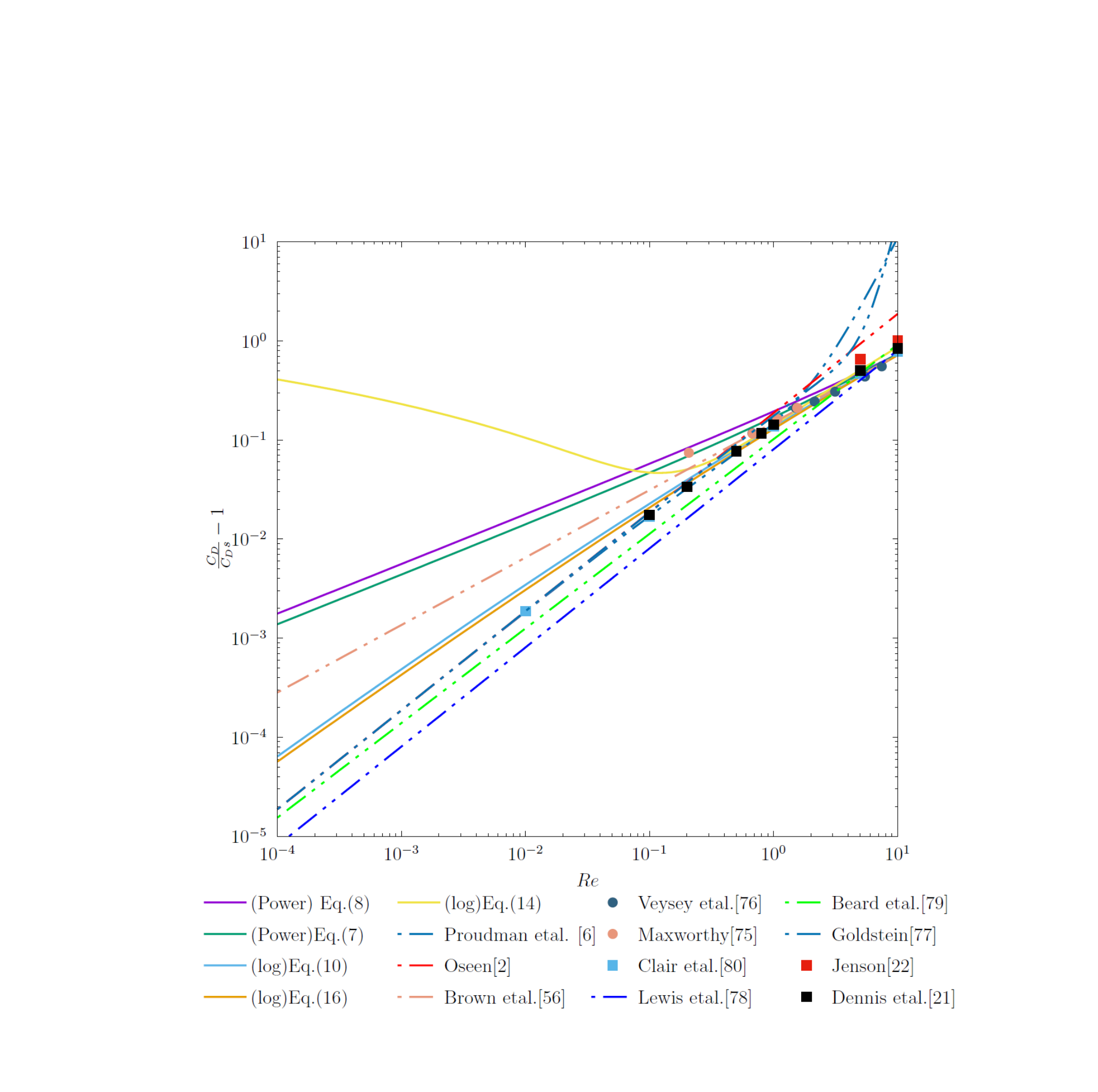}
\end{center}
\caption{Comparison between the $C_D$ predictions in the low $Re$ limit by Eq.(\ref{eq7}), Eq.(\ref{eq8}), Eq.(\ref{eq10}),Eq.(\ref{eq14}), and Eq.(\ref{16extra}), and  different sources from the literature for  low $Re$ regime.\  Circles represents experiments, and squares represents numerical simulations.   }
\label{Fig111}
\end{figure}

The predictions for the variation of $\dfrac{C_D}{C_{Ds}}-1$  against $Re$ from our models and numerous sources from  literature are shown in Figure \ref{Fig111}.\ In the range of $Re$ $10^{-1}$ to $10$, which is within the range of the  training data, all our derived equations, plus the Brown and Lawler \cite{brown2003sphere} correlation, follow with reasonable accuracy the experimental results of Maxworthy \cite{maxworthy1965accurate}, and Veysey and Goldenfeld \cite{john2007simple}, in addition to the numerical results of Jenson \cite{jenson1959viscous}, and Dennis and Walker \cite{dennis1971calculation}.\ In the same $Re$ range, the analytically derived equations of Proudman and Pearson \cite{proudman1957expansions}, Goldstein \cite{goldstein1965aeronautical}, and Oseen \cite{oseen1910uber} deviate from experimental, and numerical results, because of their limited applicability range.\\

Next we turn to the $Re$ range between $10^{-4}$ to $10^{-1}$, which is beyond the training data range.\ In this flow regime, the logarithm-based equations(\ref{eq10}) and 
(\ref{16extra}) follow closely the analytical results of \cite{{proudman1957expansions}, {goldstein1965aeronautical},{oseen1910uber}}, and the semi-empirical and empirical correlations of \cite{lewis1949some,beard1969determination}, and the numerical simulations of \cite{le1970numerical}.\, On the contrary, the power-based equations Eq.(\ref{eq8}), and Eq.(\ref{eq7}), as well the Brown Lawler \cite{brown2003sphere} correlation, divert significantly from the analytical, experimental, and numerical data.\ For example  the relative difference for the prediction of $\dfrac{C_D}{C_{Ds}}-1$ between Eq.(\ref{eq10}) and the analytical solution of Proudman and Pearson \cite{proudman1957expansions} is 240\% at $Re$ = $10^{-4}$. \ At the same conditions, the relative difference between the Brown and Lawler \citep{brown2003sphere} correlation and Proudman and Pearson \cite{proudman1957expansions} is 1410\%, which is significantly higher than the error generated by both logarithmic equations.\ The five times increase in the accuracy of the logarithmic based equations Eqs(\ref{eq10},\ref{16extra}) compared to the power-based equations Eqs(\ref{eq7},\ref{eq8})  suggests that  the  logarithmic equations contain terms that describe the physical reality better.\ Another interesting aspect of the results of Figure \ref{Fig111} is that it shows that we can improve the accuracy of  machine learning models for the same training data set, by using previous physical knowledge about the problem at hand.\ 
The observation from Figure \ref{Fig111} is similar to our observations for the Maclaurin expansion of the $\sin(x)$ function in the Appendix A.\ In both cases,   only  equations that have similar terms to the actual representation of a  function, or the  physical law  that they are approximating, generalize well beyond their training data.\ The results from Figure \ref{Fig111}, show that the popular power-based representation of $C_D$ fails to extrapolate beyond the range of $Re$ that is used for its training, which indicates that the power-based  representation may have only  been a convenient  mathematical fit, rather than having  physical significance.\ Finally, we want to explain why Eq.(\ref{eq14}) diverges even though it consists of logarithmic terms similar to the previous two.\ The reason for the divergence is the  $\dfrac{a_3\log^2(Re)}{Re}$ term which increases its value as the value of $Re$ decreases.\ This term can be considered an overfitting parameter, which it is easy to spot , due to the interpretable nature of the results of  symbolic regression.
\subsubsection{Critical flow regime}    
The critical flow regime is less well investigated, neither experimentally or numerically, compared to the subcritical or lower $Re$ regimes.\ There are not  any analytical approximations for  $C_D$ in the critical flow regime. Even  direct numerical simulations (DNS) are limited to the onset of the subcritical flow regime at $Re$ = $10^4$ \cite{beratlis2019origin}.\ Current computational fluid dynamics (CFD) simulations  that deal with the critical flow regime use different approximations to deal  with turbulence. Constantinescu et al. \cite{constantinescu2002detached}  use  Detached-Eddy-Simulations (DES), which is a hybrid method that combines Reynolds-Averaged Navier-Stokes (RANS), and Large Eddy Simulations (LES). \ Nakhostin and Gilijahus \cite{nakhostin2019investigation}  used RANS turbulence models for their simulations, and  Muto et al.\citep{muto2012negative}  used  Large Eddy Simulations coupled with the a subgrid-scale turbulence model.\ The most extensive numerical simulations in the critical and supercritical regime have been  conducted by Geier et al. 
 \cite{geier2017parametrization}  using a Cumulant Lattice Boltzmann method, and they do not use any turbulence models.\ Their high fidelity model uses a fourth-order accurate diffusion approach, suitable for low viscosity  of high $Re$ flows.\ The accuracy of the Cumulant Lattice Boltzmann depends on the optimization of  its parameters.\ The authors used a spectrum of three different mesh grid schemes, namely a course one with $40\times 10^6$ nodes, a medium one with $75\times 10^6$ nodes, and a fine grid mesh with $133\times 10^6$ nodes. \\

\begin{figure}[p]
\begin{center}
\includegraphics [scale=0.8, trim = 0 150 0 0,clip]{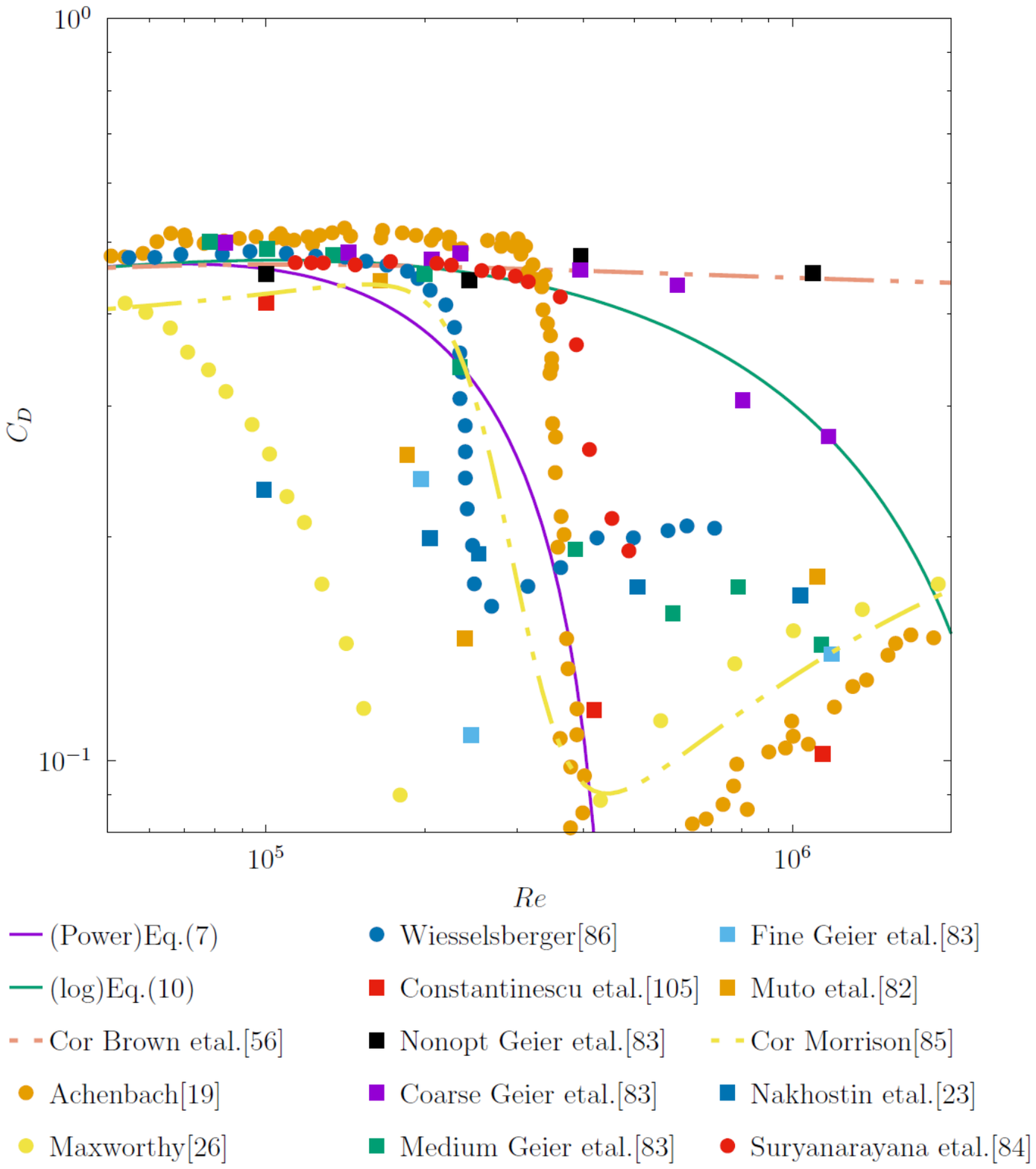}
\end{center}
\caption{Comparison between the $C_D$ predictions by Eq.(\ref{eq7}),Eq.(\ref{eq10}), and different sources in  the high $Re$ regime where the drag crisis occurs.\  Circles represents experiments, and squares represents numerical simulations.}
\label{Fig333}
\end{figure}

Figure \ref{Fig333} explores the performance of the power-based Eq.(\ref{eq7}) and logarithm-based Eq.(\ref{eq10}) in  the subcritical,   critical,  and supercritical flow regimes, and compares their performance against experimental and numerical results.\ The training data for Eq.(\ref{eq7}) and Eq.(\ref{eq10}) was limited to $Re$ up to $2\times10^5$.\ There is a significant discrepancy between the different experimental results,  for different reasons, such as the turbulence intensity the positions of the sensors around the sphere \cite{batchelor2000introduction}.  Eq.(\ref{eq7}) follows the anticipated trend in the critical flow regime in which the $C_D$ is decreasing with increasing  $Re$.\ Note that on the contrary, the value of     $C_D$ from the correlation of Brown and 
Lawler \cite{brown2003sphere}  stays constant for $Re$ values higher than $10^4$.\ The onset of the critical flow regime for the power-based equation Eq.(\ref{eq7}) starts at approximately   $Re\approx 10^5$, earlier than most  experimental and numerical results, except the experimental data of Maxworthy  \cite{maxworthy1969experiments}, in which the critical flow regime starts at much lower $ Re$.\ At approximately  $Re = 3\times 10^5$ Eq.(\ref{eq7}) drops to zero, and its values resembles  the experimental values of Achenbach \cite{achenbach1972experiments}.\ The drop of Eq.(\ref{eq7}) to zero at $Re$ = $3\times 10^5$ was already predicted algebraically in the previous section, and Eq.(\ref{eq7}) is the first in literature that predicts  with good accuracy  the value of $Re_{cr}$  reported by the  experiments of  Achenbach \cite{achenbach1972experiments}.\ From Figure \ref{Fig333},  we can see that even the high fidelity simulations of Gerier \cite{geier2017parametrization} with fine grid failed to predict $Re_{cr}$ since they failed to resolve the Kolmogorov length scale at such high $Re$.\ The numerical results for the medium grid scheme of Gerier et al. \cite{geier2017parametrization} are close to the predictions of the Eq.(\ref{eq7}) for $Re$ until the critical Reynolds number.\ The logarithmic based equation (\ref{eq10})  predicts the onset of the critical flow regime  with great accuracy since it follows the $C_D$ values from the experiments of Suryanarayana et al.
 \citep{suryanarayana1993bluff}, and Achenbach \cite{achenbach1972experiments} from $Re$ = $5\times10^4$ to about $3\times 10^5$.\ Eq.(\ref{eq10}) does not does not drop to zero at $Re_{cr}$ as Eq.(\ref{eq7}), however it follows very closely the high fidelity numerical results of Gerier et al. \citep{geier2017parametrization}, for the coarse grid case for  $Re$ up to $10^6$.\ This shows that Eq.(\ref{eq10}) follows an approximately physical reality for $Re$ up to  $10^6$, since the results of Gerier et al.
  \citep{geier2017parametrization} are generated by solving an approximate form of the Navier-Stokes equations. \ Both Eq.(\ref{eq7}) and Eq.(\ref{eq10}) fail to predict the increase of  $C_D$  after the end of the critical flow regime, and the start of the supercritical flow regime at which the boundary layer attached at the surface of the sphere changes from being partly laminar  to being fully turbulent.\ This failure is attributed to the fact the training data used to obtain Eq.(\ref{eq7}), and Eq.(\ref{eq10}) are far from the critical flow regime. Predicting  $C_D$ for the critical flow regime is  difficult even for high fidelity solvers.\ For example, the non-optimized (Nonopt) solver of 
  Gerier et al. \cite{geier2017parametrization} failed to predict the drag crisis. Instead, it predicts that $C_D$ does not change with $Re$, similar to what the correlation of Brown and Lawler \cite{brown2003sphere} predicts.\ Eq.(\ref{eq7}), and Eq.(\ref{eq10}) perform better in the critical regime than the fitting correlation of Morrison \cite{morrison2013introduction} which is a result of fitting experimental data from the literature.\  Another interesting observation is that the rate of change of  $C_D$ with  $Re$ in the critical flow regime, for both Eq.(\ref{eq7}) and Eq.(\ref{eq10}), follows the smooth trend similar to the experiments of Maxworthy \cite{maxworthy1969experiments} and the high fidelity simulations of 
  Geier et al. \cite{geier2017parametrization}, rather than the sharp nearly discontinuous  change of  $C_D$ observed in the experiments of \cite{achenbach1972experiments, suryanarayana1993bluff, wieselsberger1922}.\\

Both Eq.(\ref{eq7}), and Eq.(\ref{eq10}) predict different stages of the critical flow regime with surprising accuracy. They are the first in  literature to make such predictions without being exposed to the critical flow regime, but only by using a limited amount of physics stored in the training data and the imposed functional forms.\ The question  may arise whether  these predictions are just a product of chance? Our short answer is no, for several reasons.\ The first reason is that the $Re$ number changes by orders of magnitude in the critical flow regime, which gives many  possibilities for the output of the predictive function, but Eq.(\ref{eq10})  predicts with nearly zero error the experimental results of 
Suryanarayana et al. \cite{suryanarayana1993bluff} concerning  the onset of the critical flow regime.\ The same applies to the $Re_{cr}$ predicted by Eq.(\ref{eq7}) compared to the experimental results of Achenbach \cite{achenbach1972experiments}.\ The second and  more supportive reason is that  symbolic regression  can generalize and predict the approximated function's unexpected behaviour, similar to the example  shown in  Appendix A about the $\sin(x)$ approximation.\ The algorithm was trained to predict the peaks; however, it also accurately predicts the existence of valleys.\  We strongly believe that  Eq.(\ref{eq10}) contains terms that approximate the fundamental physical law that  $C_D$ is following, which is why it managed to generalize both  the Stokes and  critical flow regimes.\ This makes the logarithmic representation of  $C_D$ a serious candidate an the analytical  mathematical formulation that governs the  variation of $C_D$  with the $Re$.\\

In summary, in this section we showed that the functional form of $C_D$ could be represented  by both powers and logarithmic functions of $Re$.\ However, the logarithmic representation conveys  the physics in a different way than the power representation, and illuminates new physical phenomena, which are beyond the reach of  current analytical, or empirical  $C_D$ formulas.\ Because of  the logarithmic equations' good generalization behaviour, especially Eq.(\ref{eq10}), such equations should not be considered as merely  fitting equations, but rather as semi-analytical equations.\ When appealing to mathematical aesthetics, our results suggest that the drag coefficient of a sphere might be well described by the form $C_D = \pi + 24/Re + f(\log Re)$, with $C_D = \pi$ at the first point of separation, occurring at a Reynolds number $Re_s$ given by the transcendental equation  $24/Re_s + f(\log Re_s) = 0$.\ Van Dyke \cite{van1975perturbation} described the appearance of logarithms in the asymptotic expansions as obscure, but it appears that these obscure entities can speak the language of fluid dynamics much better than  powers.\ A similar situation exists in the field of turbulence, especially regarding  channel flow, where there is an open debate in the scientific community whether   power or logarithmic expansions  bests describe the velocity at the wall in certain flow regimes \cite{schultz2013reynolds}.\ Note that the logarithmic dependence of the drag coefficient $C_D$ also exists for geometries different than a sphere such as spherocylinders, and prolate spheroids, as shown by our previous work \cite{el2019solving}.

\subsection{Nusselt number $Nu$}
In this section, we will explore the possibility of a logarithmic dependence of the Nusselt number  $Nu$ on the Peclet number  $Pe$ and  Reynolds number $Re$.\ For this purpose we will create a data set of 26,796 points from the Whitaker \cite{whitaker1972forced} correlation Eq.(\ref{eq5}) for   $Pr$ in the ranging   from  0.74  to 7.0, and  $Re$ in the range of 10$^{-1}$ to $10^4$.\ We will start with the simplest assumption by allowing the  symbolic regression algorithm to guess about the dependency of $Nu$ on $Re$, $Pr$ and/or $Pe$, through the following initial function:
\begin{equation} \label{eq20}
Nu = f(Re, Pr, Pe)
\end{equation} 
The resulting $Nu$ correlation is the following: 
\begin{equation} \label{eq21}
Nu = a_1 + a_2\sqrt{Pe} + a_3\sqrt{Re}\sqrt{a_4 + a_5\sqrt{Pe}} + a_6Pe + a_7Re
\end{equation}
The coefficients are listed in Table \ref{table6}.\ Most equations that the algorithm produces show that  $Nu$ is a function of $Re$ and $Pe$, and excludes the explicit dependence  on $Pr$.\ This is  different from the source of our data (the Whitaker correlation Eq.(\ref{eq5})), which explicitly depends on $Pr$ and $Re$.\ Even when  we used a  substantial amount of data, the algorithm failed to predict the exact structure of the Whitaker correlation \cite{whitaker1972forced}. \ The recent investigation of Udrescu and Tegmark \cite{udrescu2019ai} showed, consistent with our results, that Eureqa failed to predict the exact functional structure of many  functions included in the Feynman lectures \cite{feynman1965feynman}.\ They attributed this failure due to the complexity of those functions, and the number of variables that they contain. \\

Examining the properties of Eq.(\ref{eq21}), we find  that as $Re \to 0$, Eq.(\ref{eq21}) reduces to $a_1 + a_2\sqrt{Pe}$,  which bears similarities with Eq.(\ref{eq4}) for the $Pe$ dependency, because for both cases the power of $Pe$ is less than one, and both equations show that even at very low $Re$  convection affects the heat transfer rate.\ This type of dependency did not exist in the Whitaker correlation Eq.(\ref{eq5}), where for  $Re \to 0$ (outside the range of validity of the Whitaker correlation)  $Nu$   converges to a value of 2.0,   corresponding to pure conduction from a single sphere. \\

 We will now  examine the full dependence of  $Nu$ on  logarithms of $Pe$, $Re$, and $Pr$.\ This structure of dependency is based on our previous knowledge of the physics of the problem of forced convection over a sphere.\ We know that for $Re \to 0$  and $Pe <1$,  $Nu$ depends on  $\log(Pe)$ \cite{acrivos1962heat} (Eq.\ref{eq3}), so there may exist an intermediate $Pe$ regime where logarithms  will play a role as well, until we reach a high $Pe$ regime where Eq.(\ref{eq4}) is dominant.\ For the high $Re$ regime we already showed that the drag  coefficient $C_D $ is a function of  logarithms of $Re$, so because of the tight relation between flow and heat transfer \cite{duan2015sphere} we expect that logarithms of $Re$ will play a role in the convective heat transfer process as well.\ The initial  function has  the following form:
\begin{multline} \label{eq24}
Nu = f(\log(Pe), Pe \log(Pe), \log^2(Pe), \log(Re), Re\log(Re),\\
 \log^2(Re), \log(Pr), Pr\log(Pr), \log^2(Pr))
\end{multline}
As initial guess  we gave equal weight to all  functional forms, to avoid any bias, toward any of the independent variables.\ The symbolic regression algorithm  found the following two correlations: 
\begin{equation} \label{eq25}
Nu = a_1 + a_2\log^2(Re)\log(Pe)Pe^{a_3} + a_4Pe^{a_5}
\end{equation}

\begin{equation} \label{eq26}
Nu = a_1+ a_2\log^2(Re) + a_3Pe^{a_4} + a_5\log^2(Re)\log(Pe)Pe^{a_6} +a_7\log(Pe)
\end{equation}

The second equation is more complex than the first.\ The coefficients of both Eq.(\ref{eq25}), and Eq.(\ref{eq26}) are  listed in Table \ref{table6}.\ Both equations posess  very interesting features.\ We will  start with Eq.(\ref{eq26}), where  the term $a_1+a_3Pe^{a_4}$ resembles closely the  approximation of Eq.(\ref{eq4}).\ The relative difference of  the $a_1$, $a_3$ coefficients and those of Eq.(\ref{eq4}) is 15\%, and 8\%, respectively.\ The relative error is remarkably small, if we take into  account that the source of the data set is coming from an empirical correlation that has an average predictive error of 30\%.  \\

 We believe that the combination of the logarithmic dependence of $Pe$ and $Re$ plays an essential role in the emergence of an asymptotic solution.\ It seems there  are very few possible ways to fit the data of \cite{whitaker1972forced} using logarithms of $Pe$ and $Re$ and one of those few is using terms similar to   Eq.(\ref{eq4}).\ Our findings show the essential role played by  previous physical knowledge of the problem in specific regimes, to help the machine learning algorithm to reach a physically meaningful result.\\
 
\begin{figure}[p]
\begin{center}
\includegraphics [scale=0.8, trim = 0 250 0 0,clip]{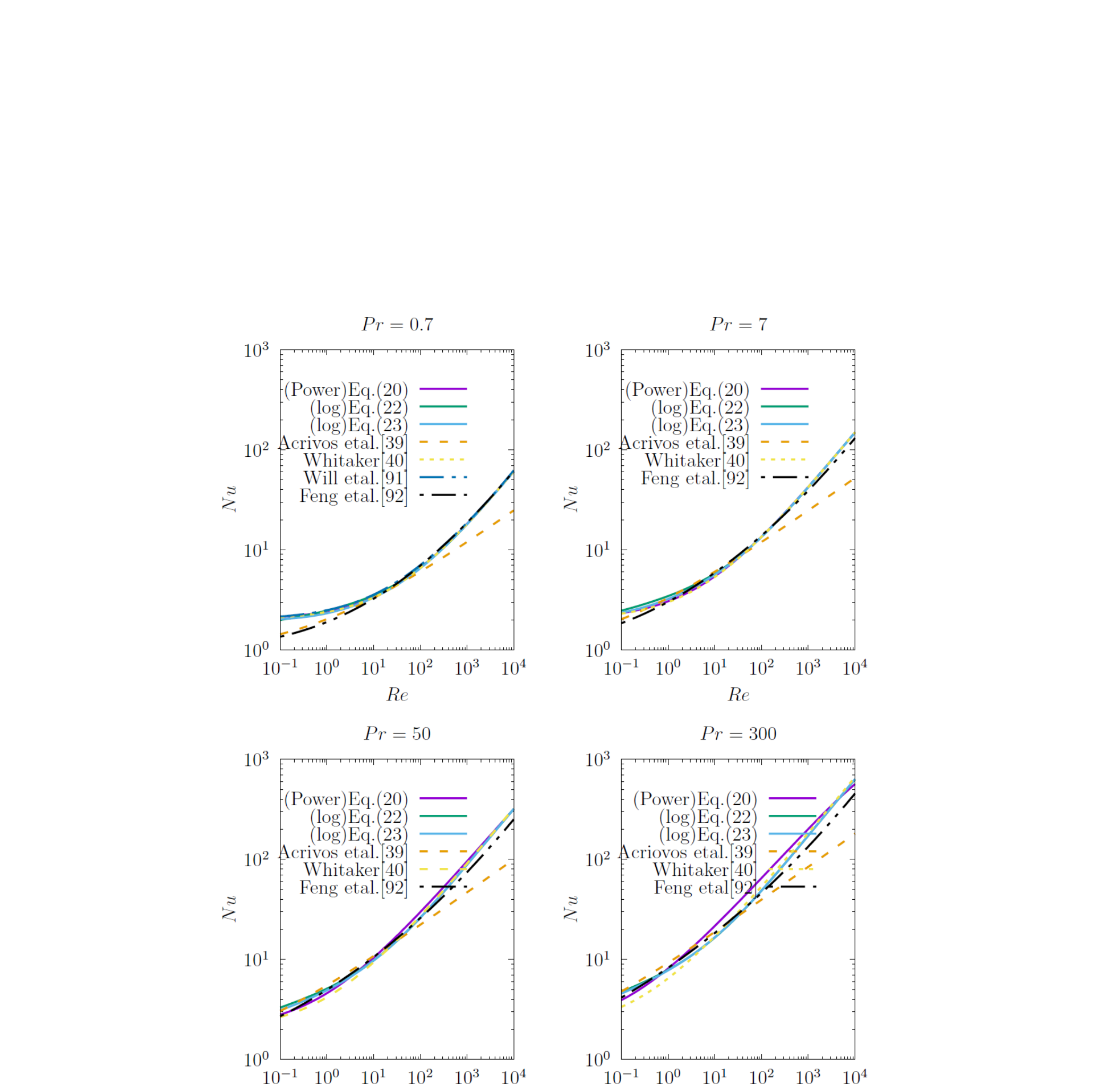}
\end{center}
\caption{Comparison  between the results of different predictor equations  for the Nusselt number $Nu$   with those  from  literature for four different Prandtl numbers $Pr$.}
\label{Fig4}
\end{figure}

The genetic algorithm predicted the asymptotic solution for the high $Pe$ (Eq.\ref{eq4}) case, rather than for low $Pe$ (Eq.\ref{eq3}), probably  because our training data  is more biased toward the high $Pe$ regime.\ Since the lowest $Re$ and $Pr$  used are 0.1 and 0.7 respectively,  the lowest  $Pe$ we used is 0.07, which lies at the boundary of the high $Pe$ regime. We could not use lower $Pe$ because  the Whitaker correlation \cite{whitaker1972forced} is based on $Re$ ranging between 3.5 
 and  $7.6\times10^4$, and $Pr$ ranging between 0.7 and 380.\ Note that we did use  the Whitaker correlation \cite{whitaker1972forced} also for lower Re, $0.1 < Re < 3.5$, to generate our training data.\ We test its validity against the experimental data of 
 Will et al.  \cite{will2017experimental} for the lowest Prandtl number  that we used, $Pr$ = 0.7, and for  $Re$ as low as 0.1, and we found that the Whitaker correlation \cite{whitaker1972forced} follows closely the results of \cite{will2017experimental}, as shown in Figure \ref{Fig4}.\ An indication that the hydrodynamics in the highly inertial regime may be governed by logarithmic terms of $Re$, is the the appearance of $\log^2(Re)$ terms both in Eq.(\ref{eq25}) and Eq.(\ref{eq26}), similar to the case of $C_D$ (see Eqs.(\ref{eq10}), (\ref{eq14}) and (\ref{16extra})). Also, the $\log^2(Re)$ terms for both  $Nu$ and $C_D$ share the same sign, and their pre-factors are of the same order of magnitude. \\

 We  compare the performance of  our predictor equations for different $Pr$, and $Re$ numbers,  in Figure \ref{Fig4}. \  We select four cases,  two of them lie within the training data set ($Pr$ = 0.7 and 7.0) that we supplied to the algorithm.\ The other two test cases ($Pr$ =50 and 300)  lie outside the training data set to test the extrapolation capabilities of our predictor equations.\ For $Pr$ = 0.7, Eqs (\ref{eq21}), (\ref{eq25}) and (\ref{eq26}) perfectly follow the Whitaker\cite{whitaker1972forced} correlation and the experimental results of Will et al.\cite{will2017experimental}.\ At high $Re$ they  also follow  the numerical results of Feng and Michaelides \cite{feng2000numerical}.\ As expected, our ecosystem of equations do not follow the asymptotic solution of Acrivos and Goddard \cite{acrivos1965asymptotic} since their solution is only valid in the  low $Re$ and high $Pe$ regime.\ For the case of $Pr$= 7.0, our ecosystem of equations predicts the evolution of $Nu$ with great accuracy.\ For the cases of $Pr$ = 50, and 300, Eqs.(\ref{eq25}) and (\ref{eq26}) predict with  great accuracy the results of the Whitaker\cite{whitaker1972forced} correlation, except in a very narrow region at  low $Re$.\ The conditions in this low $Re$ - high $Pr$ regime are applicable to  the asymptotic   solution of Acrivos and Goddard \cite{acrivos1965asymptotic}.\ This is why  the whole ecosystem of our equations deviate from the results of the Whitaker\cite{whitaker1972forced} correlation, and follow by different degrees of accuracy  the asymptotic solution of Acrivos and Goddard \cite{acrivos1965asymptotic}, Eq.(\ref{eq4}).\ All of our equations are functions of $Pe$ and $Re$.\ However, for  low $Re$ the $Nu$ correlations switch to a   dependency on $Pe$ only, which is consistent with the physics of Eq.(\ref{eq3}) and (\ref{eq4}).\\

The above  shows that  symbolic regression can find an asymptotic solution by using previous physical knowledge, rather than depending completely on the training data set.\  Feeding machine learning algorithms  previous physical knowledge for the problem that they try to optimize, increases substantially  the probability of better extrapolation predictions.\ For further discussion on how to implement previous knowledge into symbolic regression, the readers is referred to our recent publication \cite{ el2019solving}. 
\section{Conclusions }
In this investigation, we  explored the possibility of a logarithmic dependence of the drag coefficient $C_D$ on the Reynolds number $Re$, and the  Nusselt number $Nu$ on  $Re$ and  Peclet number $Pe$, inspired by  asymptotic solutions for  creeping flow conditions.\ We used a symbolic regression machine learning algorithm, and our training data are based on experiments, and data from well-known empirical correlations available in the literature.   We can make  the following conclusions : 
\begin{itemize}
\item The drag coefficient $C_D$ can be expressed as a function of powers in  $\log(Re)$, partially fulfilling the Proudman and Pearson \cite{proudman1957expansions} conjecture \textbf{P\&P}.
\item If an expansion in terms of $\log(Re)$ is made for the drag coefficient $C_D$, the value of the $Re$ at which all the $Re$ dependent terms go to zero is closely resembling the $Re$ at the first emergence  of  laminar separation, as  predicted analytically by Proudman and Pearson 
 \cite{proudman1957expansions}.  
 \item The logarithmic dependence of $C_D$ on $Re$ is found independently, without any prior knowledge, by the symbolic regression algorithm. 
\item The logarithmic based Eq.(\ref{eq10}) can generalize in both low, and high $Re$ regimes. In the high $Re$ regime Eq.(\ref{eq10}) can predict the drag crisis, its results  closely following  experimental, and numerical predictions from  literature.

\item Since Eq.(\ref{eq14})is derived from the experimental data of Brown and Lawler \cite{brown2003sphere}, the appearance of the logarithmic terms in  $C_D$ equations is independent of the  correlation that is  used as a source of the training data. 
 
 \item The Nusselt number of a single sphere depends on  logarithms of $Re$,  $Pe$, as well as  powers of $Pe$.   
 \item  If logarithmic functions of  $Re$ and $Pe$ are used as initial functions for the symbolic regression algorithm, the  algorithm produces with high accuracy the asymptotic solution derived by Acrivos and Goddard \cite{acrivos1965asymptotic} from the matched asymptotic method, in the low $Re$ and high $Pe$ regime. Interestingly, the training data that we used  does not follow the asymptotic solution of  Acrivos and Goddard \cite{acrivos1965asymptotic}.
\item There is a connection between the appearance of the logarithmic terms in both $C_D$, and $Nu$ expressions, and the ability of those expressions to generalize outside the training data range. This connection makes the logarithmic representation a strong candidate for the functional form of $C_D$ and $Nu$ that could result from solving the Navier-Stokes equations analytically for the problem of flow over a single sphere at high $Re$, and be a result of a generalized fluid mechanics theory that applies to both low and high $Re$ regimes.
\end{itemize}
 
The bigger picture of our results is that, although our method cannot give answers as rigid mathematical proofs, it is highly probably that if one day we manage to solve in a closed form the Navier-Stokes equations, combined with a heat equation around a sphere, this solution will be expressed in terms of logarithms rather than powers.\ The logarithmic terms that symbolic regression found are related to the velocity and pressure fields around the sphere. Symbolic regression is an excellent candidate to further investigate the functional form of these fields, and we intend to conduct a future study toward this goal.\ Finally, we note that the machine learning framework that we developed is general and can be used in different scientific disciplines with the condition that experimental and numerical data exists, plus the availability of some limited analytical solutions.

 \section*{Acknowledgements}
The first author  thanks Dimitra Damianidou for the enlightening discussions about the subject. The authors thank Lorenzo Botto for the discussion about  matched asymptotic methods. Finally, the authors thank the European Research Council for its financial support under its consolidator grant scheme, contract no. 615096 (NonSphereFlow).

\renewcommand{\theequation}{A.\arabic{equation}}
 
\setcounter{equation}{0}
\section*{Appendix A} \label{A}
\section*{Maclaurin expansion of Sin function}
A well-known result of applied mathematics is the representation of  continuous functions by the  Taylor expansion\cite{taylor1717methodus}: 
\begin{equation}\label{eq.1s}
f(x)= \sum_{n=0}^\infty \frac{f^n(a)(x-a)^n}{n!}
\end{equation}
When $a$ = 0, the Taylor series reduces to the Maclaurin series.\ The following expansion gives the Maclaurin series for $\sin(x)$:
\begin{equation}\label{eq.2s}
\sin(x) = \sum_{n=0}^\infty\frac{(-1)^n}{(2n+1)!}x^{2n+1} = x-\frac{x^3}{3!}+\frac{x^5}{5!}-\frac{x^7}{7!}+...
\end{equation}
One of the reasons we choose the $\sin(x)$ function as our test case for the symbolic regression algorithm is its non-monotonic nature, specifically its  transition from an increasing to a decreasing function. This feature will help us assess the generalization behaviour of the algorithm.\ We generated 5000 uniform training points in the range  [0,$\dfrac{\pi}{2}$].\ We selected this specific range because we wanted to feed the algorithm  only the monotonically increasing part of the $\sin(x)$ function, and see if it can generalize, and predict the decreasing part of the function between [$\dfrac{\pi}{2},\pi$].\ The algorithm does not possess any prior knowledge of the $\sin(x)$ function and starts by assuming the most primitive initial function for the symbolic regression algorithm:
\begin{equation}\label{eq.3s}
y = f(x)
\end{equation} 
 The symbolic regression algorithm suggested many equations, including the following two:
\begin{equation}\label{eq.4s}
y(x) =  a_1x-a_2x^3+a_3x^5-a_4x^7
\end{equation} 
\begin{equation}\label{eq.5s}
y(x) = a_1x-a_2x^3+a_4x^4+a_5x^5
\end{equation}
The values of the coefficients of Eq.(\ref{eq.4s}), and Eq.(\ref{eq.5s}) are listed in Table \ref{table1s}.\ Eq.(\ref{eq.4s}) contains the first four terms of the Maclaurin series for the $\sin(x)$ function.\ Although this may seem to be trivial, to the best of our knowledge this is the first time that a machine-learning algorithm managed to derive a Taylor or a Maclaurin series out of pure data.\ For the derivation of any Taylor series of a function we need to use the  calculus  invented simultaneously by Newton \cite{newton1833philosophiae} and  Leibniz \cite{leibniz1684}.\\

First, we want to illustrate the effect of the different terms of Eq.(\ref{eq.4s}) on its accuracy and generalization, as shown in Figure \ref{Fig1s}.\ For the [$0,\dfrac{\pi}{2}$] domain, except for the first linear term, regardless of the number of terms we add, the  decreasing nature of  $\sin(x)$  for $x>\dfrac{\pi}{2}$ is predicted. \ Adding more terms increases the accuracy.\ While the first three terms are enough to predict with great accuracy the training data, the fourth term plays a significant role for values of $x > \dfrac{\pi}{2}$ which is beyond the range of the training data.\ We chose Eq.(\ref{eq.4s}) not only because of its accuracy but due to its resemblance of the Maclaurin series, thus our selection is based on our own previous knowledge.\ What is missing is a generalization theorem which can tell us about the generalization behaviour of a specific machine learning algorithm, trained at a specific range of data. \ Without this theorem, we will always be hesitant to use machine learning predictions beyond their training range, specifically when dealing with problems for which we have minimal knowledge about the behaviour outside the training range.\ 

Finally, we want to compare the performance of the symbolic regression algorithm with other popular machine algorithms in  literature, such as polynomial regression and artificial neural networks (ANN) for the same $\sin(x)$ case.\  Polynomial regression may be considered as one of the oldest machine learning algorithms  \cite{brunton2019data}, inspired by Legendre and Gauss's works, and  implemented in a robust algorithm by Gregonne in 1815 \cite{stigler1974gergonne}.\ Polynomial regression  is the most appropriate ``traditional'' regression method to arrive at polynomials such as the Maclaurin series.\ In polynomial regression, the structure of the fitting equation  and the degree of the polynomial are predefined.\ For our case we will use two different polynomials one with a degree of $n$ = 3, and other one with $n$ = 7.\ We use the same training data set that we used for the symbolic regression, and for implementation, we will use the  Polyfit function from the open-source  Numpy library written in python \cite{numpy}.\ The main output of the algorithm is the coefficients  of the following equation: 
\begin{equation}\label{eq.6s}
y(x) = a_0+a_1x+... a_nx^n
\end{equation}  
The coefficients for the two polynomials that we used are listed in Table \ref{table2s}.\\

\ We selected the artificial neural network because it is considered as a universal function approximators \cite{cybenko1989approximation,hornik1991approximation}, but also because it does not need any prior knowledge about the structure of the equation to best fit the training data, similar to the symbolic regression algorithm.\ Contrary to symbolic regression, the product of a neural network approach is not a function but the trained neural network itself.\ We will use a  feed-forward deep neural network, with eight hidden layers. The first hidden layer consists of 64 neurons, while, the remaining hidden layers contain 32 neurons, and finally an output layer containing a single neuron \cite{brunton2019data}.\ In each hidden layer we use the Relu activation function, and also we apply L2 regularization to avoid overfitting.\ The algorithm minimizes the mean square  difference between the predicted and training data,  using a  gradient descent algorithm. \ We use the open-source library TensorFlow \cite{abadi2016tensorflow} to implement the artificial neural network framework. For training, we use 40,000 training points, which is a  much higher volume compared to the other two algorithms, because deep neural networks require a large amount of data to be trained appropriately\cite{brunton2020machine}.\\

A comparison between the performance of the  three  algorithms is shown in Figure \ref{Fig2s}.\ Symbolic regression and polynomial regression were the only algorithms that predict the peaks and valleys of the $\sin(x)$ function within the range of [-$\pi$,$\pi$].\ This success can be attributed to the fact that both algorithms represent the $\sin(x)$ function as a  polynomial.\ For the case of the symbolic regression, it discovered the polynomial representation by itself.\ On the contrary, the ANN failed to generalize beyond the training data.\ We hoped that by making the network deeper, we could help the network extract sufficient features from the training data, and generalize.\ However, what we observe is that the ANN memorizes the training data instead of generalizing it.\ For example for $x$ $>$ $\dfrac{\pi}{2}$ the output of the ANN is always a constant value of one, which is the  value of $\sin(\dfrac{\pi}{2})$, and  for     
$x<0$ the output of the ANN is always a constant value of  zero, which is the value of $\sin(0)$.\ This type of  memorization by an  ANN is also observed in several other studies such as \cite{zhang2016understanding}.\ Also, the work of Kim et al. \cite{kim2020integration} showed that if  feed-forward ANN is integrated  with symbolic regression, one obtains a better generalization behaviour compared to pure ANN.\ Another interesting observation  is that despite the fact that both symbolic regression and ANN optimize the mean  square difference, they come up with totally different generalization behaviour.\\

This Appendix A showed that symbolic regression can generalize beyond the training data, and can predict a  change in the original function occurring beyond the training range. This shows the usefulness of using interpretable machine learning results, as recommended by \cite{rudin2019stop}, and it helps us understand the output function behaviour within and beyond the training range.
 \bibliography{22} 
\bibliographystyle{nature}

\begin{table}[p]
\begin{center}
\begin{tabu} to 0.8\textwidth { | X[c] | X[c] | X[c] | X[c] | }
 \hline
 Coefficients  & Eq.(\ref{eq7}) & Eq.(\ref{eq8})& Eq.(\ref{eq12}) \\
 \hline
 $a_1$  & 0.251  & 0.412 &0.505 \\
 $a_2$  & 23.620  & 23.311&23.224 \\
 $a_3$  & 0.001  & 4.119&2.762  \\
 $a_4$  & 3.255  & - & - \\
 $a_5$  & 49.291  & - & - \\
 $a_6$  & 97.537 & -  & -\\
 $a_7$  & -$2.709\times 10^{-6}$  & -  &-\\
 \hline
\end{tabu}
\end{center} 
\caption{ Coefficients for Eq.(\ref{eq7})  Eq.(\ref{eq8}), and Eq.(\ref{eq12}) }
\label{table1}
\end{table}
\begin{table}[p]
\begin{center}
\begin{tabu} to 0.8\textwidth { | X[c] | X[c] | X[c]  | X[c] | }
 \hline
 Coefficients  & Ref\cite{brauer1972stromungswiderstand} & Ref\cite{holzer2008new} & Ref \cite{abraham1970functional} \\
 \hline
 $a_1$  & 2.9\%  & -1.94\% & 29.01\%\% \\
 $a_2$  & -2.95\%  & -2.95\% & -2.87\%  \\
 $a_3$  & 2.88\%  & 27.16\%  & -28.40\%\\
 
 \hline
\end{tabu}
\end{center} 
\caption{ Relative difference in the values of coefficients of Eq.(\ref{eq8}) to that of Brauer and Mewes \cite{brauer1972stromungswiderstand},  Holzer and Sommerfeld \cite{holzer2008new}, and Abraham\cite{abraham1970functional}. }
\label{table2}
\end{table}
\begin{table}[p]
\begin{center}
\begin{tabu} to 0.8\textwidth { | X[c] | X[c] | X[c] |  }
 \hline
 Coefficients  & Eq.(\ref{eq10}) & Eq.(\ref{eq14}) \\
 \hline
 $a_1$  & 3.286 & 3.272  \\
 $a_2$  & 24.205  & 23.26  \\
 $a_3$  & -0.818  & 0.112 \\
 $a_4$  &  0.064 & -0.652 \\
 $a_5$  & -0.000107  & 0.035 \\
  \hline
\end{tabu}
\end{center} 
\caption{ Coefficients for Eq.(\ref{eq10}) and Eq.(\ref{eq14}) }
\label{table3}
\end{table}
\begin{table}[p]
\begin{center}
\begin{tabu} to 0.8\textwidth { | X[c] | X[c] |}
 \hline
 Coefficients  & Eq.(\ref{16extra})  \\
 \hline
 $a_1$  & 3.140  \\
 $a_2$  & 24.270   \\
 $a_3$  & -0.716   \\
 $a_4$  &  0.047   \\
\hline
\end{tabu}
\end{center} 
\caption{ Coefficients for Eq.(\ref{16extra}) }
\label{table3extra}
\end{table}

\begin{table}[p]
\begin{center}
\begin{tabu} to 0.8\textwidth { | X[c] | X[c] | X[c] | X[c] | }
 \hline
 Coefficients  & Eq.(\ref{eq21}) & Eq.(\ref{eq25})& Eq.(\ref{eq26}) \\
 \hline
 $a_1$  & 2.0  &1.582  & 1.063 \\
 $a_2$  &  0.343 &0.003 & 0.0067 \\
 $a_3$  & 0.0454  &0.326 &  1.351\\
 $a_4$  &  9.341 & 1.0 & 0.299 \\
 $a_5$  & 1.0  & 0.322 &  0.0028\\
 $a_6$  & $-7.0\times10^{-5}$ & - &0.332 \\
 $a_7$  & -0.00131 &  - & -0.128 \\
  \hline
\end{tabu}
\end{center} 
\caption{ Coefficients for Eq.(\ref{eq21}), Eq.(\ref{eq25}), and Eq.(\ref{eq26})}
\label{table6}
\end{table}
\newpage
\renewcommand{\thetable}{A.\arabic{table}}    
\setcounter{table}{0}
\begin{table}[p]
\begin{center}
\begin{tabu} to 0.8\textwidth { | X[c] | X[c] | X[c] |  }
 \hline
 Coefficients  & Eq.(\ref{eq.4s}) & Eq.(\ref{eq.5s}) \\
 \hline
 $a_1$  & 0.9999 & 1.0001  \\
 $a_2$  & 0.1665  & 0.1682  \\
 $a_3$  & 0.00826  & 0.0031 \\
 $a_4$  & 0.000173 & 0.0065 \\
   \hline
\end{tabu}
\end{center} 
\caption{ Coefficients for Eq.(\ref{eq.4s}) and Eq.(\ref{eq.5s}) }
\label{table1s}
\end{table}
\begin{table}[p]
\begin{center}
\begin{tabu} to 0.8\textwidth { | X[c] | X[c] | X[c] |  }
 \hline
 Coefficients  & $n$ = 3 & $n$ = 7 \\
 \hline
 $a_0$  & -0.002 & -4.70$\times 10^{-8}$  \\
 $a_1$  & 1.027  & 1.0  \\
 $a_2$  & -0.069  & -2.339$\times 10^{-5}$ \\
 $a_3$  & -0.138& -0.166 \\
 $a_4$	&-& -2.45$\times 10^{-4}$\\
 $a_5$	&-& 0.008\\
 $a_6$ 	&-& -2.046$\times 10^{-4}$\\
 $a_7$ 	&-& -1.377$\times 10^{-4}$\\
   \hline
\end{tabu}
\end{center} 
\caption{ Coefficients of polynomials of degree $n$ = 3,and 7 }
\label{table2s}
\end{table}

\renewcommand\thefigure{A.\arabic{figure}}    
\setcounter{figure}{0}
\begin{figure}[p]
\begin{center}
\includegraphics [scale=0.8, trim = 0 300 0 0,clip]{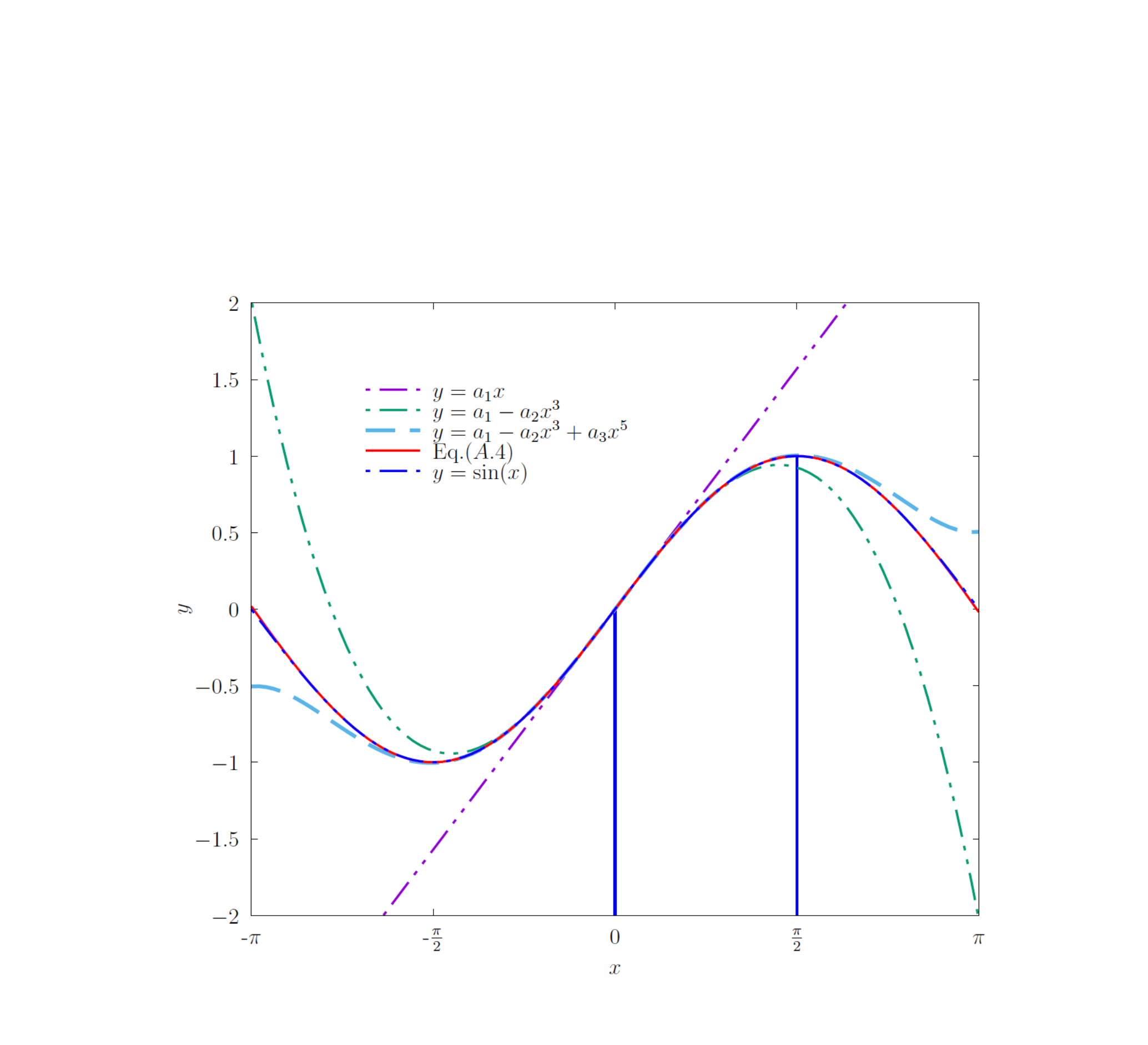}
\end{center}
\caption{The influence of different terms of Eq.(\ref{eq.4s}) on its variation with $x$. Blue bars indicate the range of training data.}
\label{Fig1s}
\end{figure}
\begin{figure}[p]
\begin{center}
\includegraphics [scale=0.8, trim = 0 300 0 0,clip]{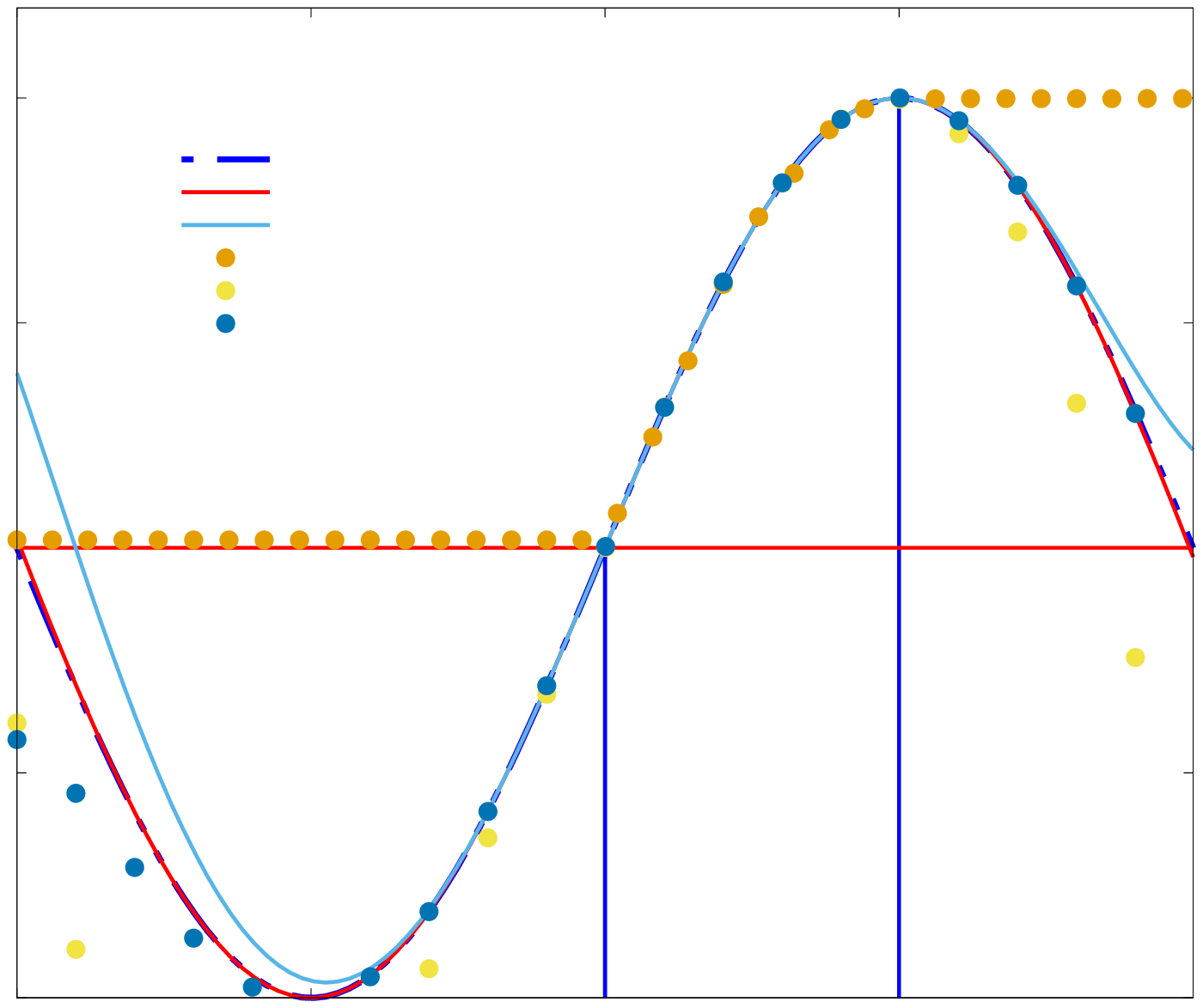}
\end{center}
\caption{Comparison between different machine learning methods for the $\sin(x)$ example. Blue bars indicate the range of training data.}
\label{Fig2s}
\end{figure}

\end{document}